\documentclass[useAMS,usenatbib,usegraphicx]{mn2e}

\usepackage{amssymb}
\newcommand{\expect}{\mathbb{E}}
\newcommand{\disp}{\mathbb{D}}
\newcommand{\FAP}{{\rm FAP}}
\newcommand{\KLIC}{{\rm KLIC}}
\newcommand{\FS}{{\rm FS}}
\newcommand{\FI}{{\rm FI}}
\newcommand{\IS}{{\rm SI}}

\title[Distinguishing between aliases]{Distinguishing between a true period and its alias,
and other tasks of model discrimination}

\author[R.V.~Baluev]{Roman V. Baluev\thanks{E-mail: roman@astro.spbu.ru}\\
Central (Pulkovo) Astronomical Observatory of Russian Academy of Sciences, Pulkovskoje sh. 65/1, St Petersburg 196140, Russia\\
Sobolev Astronomical Institute, St Petersburg State University, Universitetskij prospekt
28, Petrodvorets, St Petersburg 198504, Russia}

\begin{document}

\date{Accepted 2012 February 21.
      in original form 2011 May 31}

\pagerange{\pageref{firstpage}--\pageref{lastpage}} \pubyear{2011}

\maketitle

\label{firstpage}

\begin{abstract}
We consider the task of distinguishing between two different alternative models that can
roughly equally explain observed time series data, mainly focusing on the period ambiguity
case (aliasing). We propose a test for checking whether the rival models are
observationally equivalent for now or they are already distinguishable. It is the Vuong
closeness test, which is based on the Kullback-Leibler Information Criterion. It is
asymptotically normal and can work (in certain sense) even in the misspecified case, when
the both proposed alternatives are actually wrong. This test is also very simple for
practical use. We apply it to several known extrasolar planetary systems and find that our
method often helps to resolve various model ambiguities emerging in astronomical practice,
but preventing us from hasty conclusions in other cases.
\end{abstract}


\section{Introduction}
The search for periodicities is one of the most basic tasks of the observational data
analysis. This task emerges in all or almost all branches of astronomy (and even not only
astronomy), which deal with observational (or experimental) data. This task is usually
solved by means of the periodogram-based approach. In this approach, one considers certain
function of a period (or frequency), which basically represents some estimation of the
power spectrum of the observed process. Such estimating function is traditionally called a
\emph{periodogram}, and there are many types of periodograms that are used in practice.
The \citet{Lomb76} -- \citet{Scargle82} periodogram is a popular choice, for instance.

The observed data are usually noisy and, consequently, such data produce noisy
periodograms. This implies that some statistical issues should be usually resolved during
the period search procedure. The first such issue is the determination of the statistical
significance of the extracted periodicities. This problem is already investigated in
advance by various authors, see e.g. \citep{Frescura08} and \citep{Baluev08a} for a recent
review.

The present work is devoted to another issue, which arises when the original data are not
spaced uniformly in time. In astronomy, it is a frequent case when the observations a
gapped in time due to several natural phenomena like the day/night cycle, moonlight
contamination, etc. It is well-known that such data produce so-called \emph{aliases}, that
is false stroboscopic periods contaminating the periodograms. Sometimes, it is even
difficult to say, which periodogram peak is the real one and which is its alias. Our paper
will focus on this situation of the alias ambiguity. The periodograms themselves are
usually easily calculated using pretty simple formulae. Recent results \citep{Baluev08a}
also allow very simple (but rigourously justified) assessing of the significance of the
extracted periods. However, there is a lack of similarly simple and rigorous methods
solving the alias discrimination task.

In a particular application, we usually have, in addition to the raw time series data,
some prior information concerning the phenomenon under research. This information often
can be used to select one of the available alternatives, or at least to retract some of
them, making the final pool of solutions more narrow. For example, in the exoplanet radial
velocity searches, a popular approach involves various dynamical stability or regularity
criteria \citep{Gozd06b,Gozd07,Gozd08}. No doubt, such methods are useful, but we would
often prefer not to bind ourselves to any external assumptions as much as possible,
allowing the data to speak for themselves. For example, in the case of exoplanet searches,
the apparent radial velocity patterns can often be caused by stellar activity effects
rather than by orbiting planets. In such a case any dynamical criteria may be unreliable,
since we cannot be sure even that a given radial velocity oscillation is indeed induced by
a really existing planet.

In other words, in this work we put a goal to find a purely statistical method of alias
discrimination, that could be maximally independent on various prior assumptions. We may
highlight a recent work by \citet{Dawson10}, where the authors propose to take into
account the phases of candidate aliases, when choosing between them. However, this
approach is not very rigorous, because it basically does not take into account any
possible noisy fluctuations of these phases, as well as of the aliases amplitudes. In
addition, this approach is not easy to apply in an automated way, since it requires an
active participation of the researcher.

The source of the lesser attention to this problem probably comes from the fact that to
distinguish aliases we need to compare two \emph{non-nested} models of the data, while
most of the traditionally used statistical criteria are usually designed to choose between
\emph{nested} models. For example, the signal detection task requires to test some
``base'' model against a more general ``base+signal'' model, which includes the original
base model as a partial case. The alias ambiguity does not infer nested models: none of
the two alternative periodicities can encompass another one as a partial case. Such
situation usually falls out of a typical handbook on mathematical statistics.
Nevertheless, we find that there are many individual statistical works dealing with tests
for non-nested hypotheses, so this statistical problem appears well-studied. Our goal in
this paper is to apply these results to the astronomical task of resolving the alias
ambiguity, and to construct the appropriate period distinguishing criteria.

The problem is also complicated because we often do not have an accurate model of the real
periodic variation. In practice, both alternative models (e.g. simple sinusoids) may be
only approximate, and our analysis cannot base on an assumption that one of them is
strictly true. Therefore, the method of the analysis should remain valid in the so-called
misspecified case, when none of the available models can mimic the true signal precisely,
but we are still interested to find out which alternative is better.

The plan of the paper is as follows. In Section~\ref{sec_formulation}, we provide a more
clear formulation of the problem, what we actually want to derive from our analysis. In
Section~\ref{sec_fundamentals}, we describe the basic formulae and ideas of the Vuong
statistical test, suitable for solving our task. In Section~\ref{sec_practice}, we provide
the final practical formulae of this test. In Section~\ref{sec_applicability}, we assess
its reliability and efficiency by means of numerical Monte Carlo simulations. Finally, in
Section~\ref{sec_exo}, we apply our methods to the analysis of the radial velocity data of
several extrasolar planetary system, which demonstrate various alias ambiguities
concerning the planetary periods or other model parameters.

\section{Formulation of the problem}
\label{sec_formulation}
From the first view point, the task can be easily formulated. Assume we have two main
peaks in the periodograms, having similar height. We have already established that the
periodic signal is significant indeed (at least the larger peak or both peaks are
significant), and our first naive question is: which peak we ought to recognise as the
true one and which should be accused as its alias? A naive question has a naive answer: we
should adopt the taller peak as a more likely candidate. Then the second question arises:
how we can ensure that such choice is statistically justified? In other words, is the
observed difference between the two peaks statistically significant? If it is not then any
attempt to choose between the two models is, in fact, no better than dropping a coin. In
such situation, we can occasionaly choose the true period, but such luck cannot be
classified as a success of the analysis, since we cannot ensure that we were so lucky
indeed. Therefore, we should make available a \emph{third}, inconclusive, outcome of the
analysis. This makes the problem qualitatively more complicated, because more types of
decisions infer more freedom for making mistakes.

In the framework of the symmetric hypotheses testing with no inconclusive decisions, we
have only two possible outcomes, and basically there is only one type of mistakes:
accepting a wrong model. In practice, the two models are often treated unequally, however.
For example, the first one may be considered as more reliable, while the second one is
admitted only hypothetically. In that case we have \emph{two} formal types of mistakes,
but the mistakes to wrongly reject the base model are naturally considered as more
dangerous. Thus, when applying the traditional statistical tests we first limit the
probability of this more dangerous ``false alarm'', and only then we try to minimize the
mistakes of the second kind, if it is possible. The two types of mistakes are
antagonistic: suppressing the frequency of false alarms leads to increased fraction of
wrong non-rejections of the null hypothesis.

With \emph{three} decisions at disposal, the following types of mistakes threaten: (I)
choosing wrong model when another one is actually better, (II) not choosing any model when
one is actually better, and (III) choosing a model when objectively none is really better.
The class (I) formally contains two subclasses: to wrongly accept the first or the second
model. We consider a symmetric case here, when none of the alternative models is a priori
preferable, so these two subclasses are qualitatively equivalent to each other and can be
united into one. The three remaining types of mistakes are already qualitatively
different. We cannot simultaneously suppress mistakes of all three types, and we cannot
say which one is most dangerous in general. It depends on the current practical
circumstances: sometimes we cannot tolerate a misclassification, or sometimes an
inconclusive answer is highly undesirable. The methods of our paper are designed to
suppress the ``pink-vision'' mistake (III), i.e. the cases when we wrongly assume that the
two rival models are distinguishable on the basis of the available observations. The
main arguments supporting this choice are:

\begin{enumerate}
\item Such approach allows for an intuitive formalisation of the ``statistically
justified'' choice. We choose any of the alternatives only after we have ensured that they
are not observationally equal. Otherwise, we acknowledge that any choice would be too
risky for now, and we should live with such ambiguity until more observations resolve it.

\item Frequently, none of the proposed models may be correct, and a more accurate model
should be constructed to replace both. This task usually cannot be fulfilled reliably
until we are ensured that the two original models can be distinguished. It is usually more
justified to seek for any theory update only after we managed all its current ambiguities.

\item Consider the mutual interaction between the three possible types of the mistakes.
For example, suppressing (III), i.e. generating more inconclusive answers, will likely
suppress (I) and favours to (II), while suppressing (II) obviously favours to both (I) and
(III). In other words, (II) is antagonistic to (I) and (III), while (I) and (III) are not
mutually antagonistic. Thus it should be technically convenient to first suppress (I) or
(III), which should be broadly equivalent.

\item Putting the third-type mistakes in the first place allows us to easily reformulate the
problem in the classical hypothesis testing framework, operating in the traditional terms
(to a certain point). Indeed, our main task is now to test the null hypothesis ``the
observed divergence between the two models is consistent with random noise'' against the
alternative ``this divergence is beyond the random noise level''.
\end{enumerate}

We still need to formalise the notion of ``observational equivalence'' of the rival
models. The problem is complicated by the absence of any guarantee that at least one of
the alternative models is indeed correct. In true, the signal may have a non-sinusoidal
shape (while, e.g., the Lomb-Scargle periodogram implicitly assumes that it is always
sinusoidal), and the data may contain some extra small periodic or non-periodic
variations. We usually want to know, which model fits the observations \emph{better},
rather than which one is \emph{true}. Therefore, our test should not rely on an assumption
that one of the proposed models is correct.

We find that the Vuong test \citep{Vuong89} suits our needs very well. This method
utilises the Kullback-Leibler Information Criterion ($\KLIC$) to define which of the rival
models is ``better''.

\section{Fundamentals of the test}
\label{sec_fundamentals}
Let us consider two rival models of the data, $\mu_1(t,\btheta_1)$ and
$\mu_2(t,\btheta_2)$ parametrised by the vectorial arguments $\btheta_1$ and $\btheta_2$.
In the period ambiguity case these models (as well as the parameter vectors) are
functionally the same for both alternative models, the only difference is due to the
different values of the frequency (which is embedded inside $\btheta_{1,2}$). Nonetheless
we still use a bit more general notation then is necessary for our aliasing case. In
general, $\btheta_1$ and $\btheta_2$ may be even unrelated to each other, having different
dimensions. Given the timings $t_i$, measurements $x_i$, and expected (known) measurement
uncertainties $\sigma_i$ (for $i=1,2,\ldots N$), we may assume (just for some
definiteness) that the probability density function of each measurement is Gaussian:
\begin{equation}
f_1(x_i|t_i,\sigma_i,\btheta_1) = \frac{1}{\sigma_i\sqrt{2\pi}}
\exp\left[-\frac{1}{2}\left(\frac{x_i-\mu_1(t_i,\btheta_1)}{\sigma_i}\right)^2\right],
\label{gauss}
\end{equation}
with a similar definition $f_2$ for the second rival model, $\mu_2$. For further
shortness, we define the pairs $z_i=\{t_i,\sigma_i\}$ and two vectors: $\bmath x$
containing all $x_i$ and $\bmath z$ containing all $z_i$ (that is, $t_i$ and $\sigma_i$).
These model probability density functions $f_{1,2}$ depend on the unknown vectors of
parameters $\btheta_{1,2}$, which can be estimated from the data using the
maximum-likelihood approach, which turns into the least-square one in the Gaussian case.
Namely, the necessary estimations of $\btheta_{1,2}$ can be defined as
\begin{equation}
\widehat\btheta_k = \arg\max_{\btheta_k\in \Theta_k} \mathcal L_k(\bmath x|\bmath z,\btheta_k) =
\arg\min_{\btheta_k \in\Theta_k} \chi^2_k(\bmath x|\bmath z,\btheta_k),
\label{est}
\end{equation}
where the functions
\begin{eqnarray}
\mathcal L_k(\bmath x|\bmath z,\btheta_k) &=& \sum_{i=1}^N f_k(x_i|z_i,\btheta_k), \nonumber\\
\chi^2_k(\bmath x|\bmath z,\btheta_k) &=& \sum_{i=1}^N \left(\frac{x_i-\mu_k(t_i,\btheta_k)}{\sigma_i}\right)^2
\end{eqnarray}
represent the appropriate likelihood functions and the chi-square functions, respectively.
Note that the likelihood function also represents the joint probability density for the
vector of the observables $\bmath x$, given fixed timings, error uncertainties, and model
parameters.

Let us now try to rigorously compare the two alternative models. \citet{Vuong89} suggests
to use the Kullback-Leibler Information Criterion ($\KLIC$) to compare the models in the
sense of their separation from the truth:
\begin{eqnarray}
\KLIC_k(\btheta_k) &=& \int \log \frac{h(x|z)}{f_k(x|z,\btheta_k)} h(x,z)\, dx dz = \nonumber\\
&=& \expect^0_{\bmath x,\bmath z} \log \frac{h(x|z)}{f_k(x|z,\btheta_k)}, \nonumber\\
\KLIC_{12}(\btheta_1,\btheta_2) &=& \KLIC_2(\btheta_2)-\KLIC_1(\btheta_1) = \nonumber\\
&=& \int \log \frac{f_1(x|z,\btheta_1)}{f_2(x|z,\btheta_2)} h(x,z)\, dx dz \nonumber\\
&=& \expect^0_{\bmath x,\bmath z} \log\frac{f_1(x|z,\btheta_1)}{f_2(x|z,\btheta_2)},
\label{KLIC}
\end{eqnarray}
where $h(x,z)$ represents the true (unknown) joint probability density of a single
measurement $x$ and of the associated quantities in $z$, and $h(x|z)$ is the respective
conditional density of $x$ given $z$. This true distribution $h$ involves, the true signal
shape and true error distribution shape, instead of the modelled ones. The quantities
$\KLIC_1$ and $\KLIC_2$ assess the separation\footnote{We say ``separation'' instead of
``distance'', since this divergence measure does not satisfy all necessary axioms of the
distance, e.g. it is not symmetric and does not satisfy the triangle inequality.}
between one of the two available statistical models of the data and the truth. The
difference between these quantities, $\KLIC_{12}$, measures the ability of the first model
to reproduce the true distribution, relatively to the second one. The definitions
in~(\ref{KLIC}) do not yet involve anything from the actual observational data, they are
defined regardless of what we observe. The symbol $\expect^0$ denotes the mathematical
expectation, taken for the true distribution $h$.

It is not required that the model error distribution shape set in the functions $f_k$
should be close to $h$. In practice, we usually have no other option than to set $f_k$ to
the Gaussian functional shape~(\ref{gauss}). Then $\KLIC_{12}$ will measure the separation
between $\mu_1$ and $\mu_2$ in the sense of their expected mean-square residuals. Such
separation measure remains quite sensible and justified even when $h(x|z)$ is
non-Gaussian. Moreover, even if we know definitely the non-Gaussian shape of $h$, we may
want to compare the models in a homogeneous manner, still using the mean-square residual
for this goal, i.e. assuming Gaussian $f_k$. We may want to avoid any binding to the true
shape of $h$ as much as possible, because it is often related only to instrumental
properties rather than to physical objects that we observe.

In general, there is no ``true'' values of $\btheta_1$ and $\btheta_2$ that we could
substitute in~(\ref{KLIC}), since both rival models $f_{1,2}$ are wrong and the true
distribution $h$ is not parametrized at all. We should substitute the so-called
``pseudo-true'' values of $\btheta_1$ and $\btheta_2$, which represent some theoretically
most suitable values of these parameters. They are defined as the points where $\KLIC_1$
or $\KLIC_2$ reaches their maximums. Since the true density $h(x,z)$ is unknown, the
pseudo-true values of $\btheta_{1,2}$ are unknown too, but the estimations~(\ref{est}) can
approximate them. Eventually, we want to test the null hypothesis $\KLIC_{12}=0$ (the two
models are equally close to the truth) against the alternative $\KLIC_{12}\neq 0$.

It is essential that in the definition~(\ref{KLIC}) the timings and error uncertainties
inside $z_i$ are treated as random quantities too, so we can speak of the joint density
$h(x,z)$. Such probabilistic interpretation of apparently non-random quantities is in fact
quite natural, and we consider it as a strength of the approach. In practice, we often do
not know the exact time of each observation in advance. This time depends on many things
that are unrelated to the astronomical problem, like the observatory's time allocation
policy and current routine circumstances, the racing between different observing
programmes, etc. This issue is also discussed in more details in Appendix~\ref{sec_rtime}.
It is important that we are not required to specify/estimate $h(x,z)$ explicitly, the test
will deal with it automatically. Also, we do not make any restrictive assumptions about
the distribution of the timings $t_i$: it is allowed to be highly non-uniform, e.g. gapped
(as it usually occurs in astronomy).

Now, it is not difficult to realize that the following normalized log-likelihood ratio
\begin{equation}
L=\frac{1}{N} \log\frac{\mathcal L_1(\bmath x|\bmath z,\widehat\btheta_1)}{\mathcal L_2(\bmath x|\bmath z,\widehat\btheta_2)} =
\frac{1}{N}\sum_{i=1}^N l_i, \quad l_i=\log \frac{f_1(x_i|z_i,\widehat\btheta_1)}{f_2(x_i|z_i,\widehat\btheta_2)}
\label{lmean}
\end{equation}
represents an empirical estimation of the $\KLIC$ divergence measure~(\ref{KLIC}), since
the averaging over $l_i$ approximates the mathematical expectation in~(\ref{KLIC}). Note
that the individual terms $l_i$ are generated by the real data, so they are automatically
averaged on the basis of the unknown true distribution $h$. This trick is rather
reminiscent of the well-known jackknife or bootstrap procedures, which also use the
original sample to eliminate the necessity to specify the unknown error distribution.

Furthermore, we can find the empirical variance of $l_i$ as
\begin{equation}
v^2 = \frac{1}{N} \sum_{i=1}^N l_i^2 -\frac{1}{N^2} \left(\sum_{i=1}^N l_i\right)^2.
\label{lvar}
\end{equation}
The uncertainty of $L$ is then equal to $v/\sqrt N$, and therefore the final Vuong
statistic represents another re-normalised log-likelihood ratio
\begin{equation}
\mathcal V = \frac{L\sqrt{N}}{v} = \frac{\sum_{i=1}^N l_i}{\sqrt{\sum_{i=1}^N l_i^2-\frac{1}{N}\left(\sum_{i=1}^N l_i\right)^2}}.
\label{Vuong}
\end{equation}
Under the null hypothesis, $\mathcal V$ behaves asymptotically (for $N\to\infty$) as a
standard normal varaible (mean zero, variance unit). This result can be used to test the
models equivalence. If $|\mathcal V|$ is too large then our null hypothesis $\KLIC_{12}=0$
is inconsistent with the data, so that the rival models are well-distinguishable and we
can safely adopt the one which offers a better likelihood (i.e., the first one if
$\mathcal V>0$, or the second one otherwise). If $\mathcal V$ is not large enough then we
have to acknowledge that for now it is still too risky to choose between the alternatives
and it is better to seek for more data before drawing any definite conclusion. Given an
observed value of $\mathcal V$, we can calculate the associated false alarm probability as
$\FAP=2\Phi(|\mathcal V|)$, where $\Phi(x)$ is the standard Gaussian tail function (i.e.,
probability for a standard normal varaible to exceed a given $x$). To reject the null
hypothesis, we need to have this $\FAP$ below some small critical value $\FAP_*$
(typically, $1\%$, $5\%$, etc.).

The formula~(\ref{Vuong}) actually refers to the case when both models have the same
number of free parameters (degrees of freedom). If it is not the case, we need to add a
minor bias correction, because models with larger number of parameters always produce
systematically better fits than models with smaller number of parameters. \citet{Vuong89}
suggests to add a Bayesian-style correction as
\begin{equation}
\mathcal V = \frac{\sum_{i=1}^N l_i - \frac{d_1-d_2}{2} \ln N}{\sqrt{\sum_{i=1}^N l_i^2-\frac{1}{N}\left(\sum_{i=1}^N l_i\right)^2}},
\label{Vuong2}
\end{equation}
with $d_1$ and $d_2$ being the numbers of free parameters in the models. This correction
tends to zero when $N\to\infty$, although may remain relatively important in practice,
especially when $d_1-d_2$ is large.\footnote{When passing to the limit $N\to\infty$, the
values of $L$ in~(\ref{lmean}) and $v$~in~(\ref{lvar}) tend to certain constant finite
values. Namely, $L$ tends to the $\KLIC$ defined in~(\ref{KLIC}), and $v$ tends to a
similar expression with mathematical expectation replaced by the variance operator.
Bearing this in mind, we easily find that the difference between~(\ref{Vuong})
and~(\ref{Vuong2}) decreases as $\ln N/\sqrt N$.} In this paper we only deal with models
having the same number of parameters, so we will use only the uncorrected
formula~(\ref{Vuong}).

Since the Vuong test is based on the usual likelihood ratio statistic, it is interesing to
compare it with the traditional methods for testing nested hypotheses, when the first
model represents a parametric subspace of the second one. This traditional problem
basically represents a degenerate case in the conditions of the non-nested hypotheses
testing, and the distribution of $\mathcal V$ is then not asymptotically normal. It is
well-known that in that case the quantity $L N$ follows a chi-square distribution (if the
first model is correct). As it is discussed in the previous paragraphs, in the non-nested
case the quantity $L\sqrt N$ is asymptotically normal. Therefore, the main difference
between the nested and non-nested situations is in the magnitude of the random scatter in
$L$, about $\sim 1/N$ or about $\sim 1/\sqrt N$.

We need to emphasize that the Voung statistic is not equivalent to the textbook likelihood
ratio in~(\ref{lmean}). The only common thing between $L$ and $\mathcal V$ is their sign.
This means that the Vuong test uses in fact the likelihood function to identify which of
the models is more likely. However, as we explained in Sect.~\ref{sec_formulation}, our
main goal was to find out whether the alternatives are distinguishable or they are not.
The likelihood ratio method is not usable for this, because its distribution in this
situation is not known in general. It is only known that the distribution of $L$ is
asymptotically normal, but its variance is severely dependent on the particular data/error
model in $f$, and should be evaluated on the task-by-task basis \citep{Cox62}. The Vuong
test involves a normalization that makes the distribution of $\mathcal V$ asymptotically
invariable with respect to the model $f$.

Another very important consequence is that the Vuong test is asymptotically insensitive to
possible model faults, in the sense that $\mathcal V$ preserves its standard normal
distribution even if the true signal and error behaviour does not really match the model
$f$. Such possible shortcomings of $f$ may have two distinct reasons: (i) wrong
assumptions about the signal shape and (ii) wrong assumptions about the measurement error
distribution.

In the first case, we try to fit the observed signal using inadequate models. In this case
our test cannot suggest any third (correct) model, this task still lies on the
researcher's shoulders. However, it can advice us whether the models at disposal are
distinguishable or equivalent. The answer to this question is often important even for
inaccurate models, because if these simpler models are indistinguishable then there is a
little chance that some other more complicated model may appear more suitable (and at
least observationally distinguishable from the original models). In this case, it is
important that both Vuong test does not assume that any of the available alternatives is
true (while, e.g., in the classic base/alternative hypothesis testing it is always assumed
that either first or at least the second model is functionally correct).

In the second case, we may assume inadequate model for the error distribution. This is not
an obstacle for our test at all: the asymptotic distribution of the statistic $\mathcal V$
remains the same for the most well-behaving distributions of the measurement errors
(possibly except for some too heavy-tailed ones that always constitute a trouble, see all
rigorous formulation in \citealt{Vuong89}). As we noted above, in practice it is even
desirable not to bind our ordering criterion ($\KLIC$) to the shape of the error
distribution, since this shape is usually related to the instrumental characteristics
rather than to the physical phenomenon under study. For example, setting $f$ to a Gaussian
bell shape~(\ref{gauss}) we always order our signal models in the sense of the mean-square
deviation. When $f$ does not match the true shape of the error distribution, the quantity
in Eq.~(\ref{lmean}) and, consequently, $\mathcal V$ are related to the so-called pseudo
maximum-likelihood tests that possess many practically useful properties of the classic
likelihood ones \citep{Gourieroux84,Baluev08b}.

Nonetheless, the \citet{Vuong89} paper contains a possibly essential requirement of the
statistical independence of the pairs $(z_i,x_i)$ for different $i$. Therefore, both the
Vuong test may yield unreliable results, if the measurement noise is correlated (not
white).

\section{Practical formulae}
\label{sec_practice}
For the Gaussian distribution~(\ref{gauss}), the Vuong statistic can be derived
from~(\ref{Vuong}), substituting
\begin{eqnarray}
\l_i = \frac{\left(\widehat\mu_1(t_i)-\widehat\mu_2(t_i)\right)}{\sigma_i^2} \cdot
\left[x_i - \frac{\widehat\mu_1(t_i)+\widehat\mu_2(t_i)}{2} \right], \nonumber\\
\widehat\mu_1(t)=\mu_1(t,\widehat\btheta_1), \qquad \widehat\mu_2(t)=\mu_2(t,\widehat\btheta_2).
\label{li}
\end{eqnarray}
These formulae can be used after substitution of the best fitting alternative models
$\widehat \mu_1$ and $\widehat \mu_2$ that we wish to compare (these models are
task-specific).

In the simplified aliasing case, we have formally the same model for the both
alternatives, which represents a sinusoidal oscillation:
\begin{equation}
\mu(t,\btheta) = a \cos(\omega t) + b \sin(\omega t), \quad \btheta=\{a,b,\omega\}.
\label{sinmod}
\end{equation}
The two alternatives are different only due to the different admissible ranges for the
frequency: for the first model, $\mu_1$, the (circular) frequency $\omega$ should be
located around the first rival frequency, $\widehat\omega_1$, and for the model $\mu_2$ it
should be around $\widehat\omega_2$. In practice, it is more convenient to treat $\mu_1$
and $\mu_2$ as different models, rather than two variants of the model~(\ref{sinmod}),
because of the strong non-linearity of the frequency parameter. The test that we have
described has asymptotic nature (for $N\to\infty$), consequently it implicitly utilise
some hidden linearisation of the models in the vicinity of the best fitting estimations.
Although the coefficients $a$ and $b$ are fully linear for all values of $\omega$, the
frequency $\omega$ itself is not globally linear and is not linearisable in the global
sense, although it is still linearasable in the local sense. Therefore, we should treat
the basic model $\mu$ in the vicinities $\omega\simeq\widehat\omega_1$ and
$\omega\simeq\widehat\omega_2$ as two separate models.

Let us write down the classical expression for the Lomb-Scargle periodogram:
\begin{eqnarray}
z(\omega) &=& \frac{1}{2}\left[
 \frac{\left(\sum \frac{x_i}{\sigma_i^2} \cos\omega(t_i-\tau)\right)^2}{\sum \frac{1}{\sigma_i^2} \cos^2 \omega (t_i-\tau)}\right.+\nonumber\\
& & \qquad + \left.\frac{\left(\sum \frac{x_i}{\sigma_i^2} \sin\omega(t_i-\tau)\right)^2}{\sum \frac{1}{\sigma_i^2} \sin^2 \omega (t_i-\tau)}
\right],\nonumber\\
\tan 2\omega\tau &=& \frac{\sum \frac{1}{\sigma_i^2} \sin 2\omega t_i}{\sum \frac{1}{\sigma_i^2} \cos 2\omega t_i}.
\label{LS}
\end{eqnarray}
Note that $\tau$ is also a function of $\omega$. As it is already well-known, the
Lomb-Scargle periodogram is directly related to the likelihood ratio statistic, or, in the
Gaussian case, to the chi-square statistic \citep{ZechKur09,Baluev08a}. In addition to the
periodogram itself, we will need the expressions for the associated best fitting
coefficients $a$ and $b$ of the original model~(\ref{sinmod}). They are given by
\begin{eqnarray}
\widehat a(\omega) &=& \frac{\sum \frac{x_i}{\sigma_i^2} \cos\omega(t_i-\tau)}{\sum \frac{1}{\sigma_i^2} \cos^2 \omega (t_i-\tau)}, \nonumber\\
\widehat b(\omega) &=& \frac{\sum \frac{x_i}{\sigma_i^2} \sin\omega(t_i-\tau)}{\sum \frac{1}{\sigma_i^2} \sin^2 \omega (t_i-\tau)},
\label{ab}
\end{eqnarray}
and the final best fitting sinusoidal model evaluates to
\begin{equation}
\widehat\mu(t,\omega) = \widehat a(\omega)\cos \omega(t-\tau) + \widehat b(\omega) \sin \omega(t-\tau).
\label{bestmod}
\end{equation}

From a preceding period analysis we should have already estimated the possible rival
frequencies $\widehat\omega_1$ and $\widehat\omega_2$, which correspond to the two largest
peaks of the periodogram. Substituting these frequencies in~(\ref{bestmod}), we obtain the
two best fitting models $\widehat\mu_1(t)$ and $\widehat\mu_2(t)$, which allow to evaluate
all quantities $l_i$ using~(\ref{li}), and then the Vuong statistic~(\ref{Vuong}). In
particular, it can be readily shown that
\begin{equation}
N L = \sum_{i=1}^N l_i = z(\widehat\omega_1)-z(\widehat\omega_2),
\label{sumli}
\end{equation}
which is not surprising, since the sum of $l_i$ represents the pure log-likelihood ratio
statistic, and the Lomb-Scargle periodogram is tied to the log-likelihood function.

This is not yet the full story. In practice, the quantities $\sigma_i$ usually are not
known with good precision, and we need to model them too. For instance, the multiplicative
model $\sigma_i^2 = \kappa/w_i$ is widely used, where the factor $\kappa$ is an extra
unknown parameter, and $w_i$ are the known statistical weights. In that case, the
quantities $l_i$ are more complicated, relatively to~(\ref{li}):
\begin{eqnarray}
l_i = \frac{N}{2}\left(\frac{\widehat\nu_2(t_i)}{\widehat\chi_2^2} - \frac{\widehat\nu_1(t_i)}{\widehat\chi_1^2}\right) +
\frac{1}{2} \log\frac{\widehat\chi_2^2}{\widehat\chi_1^2}, \nonumber\\
\widehat\nu_k(t) = w(t) (x(t)-\widehat\mu_k(t))^2, \quad \widehat\chi_k^2 = \sum_{i=1}^N \widehat\nu_k(t_i)
\end{eqnarray}
In the case of the sine curve~(\ref{sinmod}) we obtain
\begin{equation}
N L = \sum_{i=1}^N l_i = \frac{N}{2} \log\frac{\widehat\chi_2^2}{\widehat\chi_1^2} =
\frac{N}{N_{\mathcal K}}\left(z_3(\widehat\omega_1)-z_3(\widehat\omega_2)\right),
\end{equation}
where the periodogram $z_3$ is defined in \citep{Baluev08a}. The number $N_{\mathcal K}$
is equal to $N-d_{\mathcal K}$, where $d_{\mathcal K}$ is the number of free parameters in
the full model with the probe sinusoid~(\ref{sinmod}), excluding the frequency parameter.
For the cases that we consider here we always have $d_{\mathcal K}=2$ (two parameters $a$
and $b$), but for more general cases, which involve an underlying variation in addition to
the periodic signal \citep[see][]{Baluev08a}, the number $d_{\mathcal K}$ may be larger.

Finally, we may use another parametrization of the measurement uncertainties, like e.g.
the so-called additive model $\sigma_i^2(p) = p+\sigma_{\mathrm{meas}, i}^2$ with a free
parameter $p$, as considered in \citep{Baluev08b}. In this case the formula for the
quantities $l_i$ will look like:
\begin{eqnarray}
l_i = \frac{1}{2}\left[ \left(\frac{x_i-\widehat\mu_2(t_i)}{\sigma_i(\widehat p_2)}\right)^2 -
\left(\frac{x_i-\widehat\mu_1(t_i)}{\sigma_i(\widehat p_1)}\right)^2 \right] +
\log \frac{\sigma_i(\widehat p_2)}{\sigma_i(\widehat p_1)}.
\end{eqnarray}
In this expression, $\widehat p_{1,2}$ represent the estimations of the parameter $p$
associated to one of the models.

In practice, we will probably use more complicated signal models than a sinusoidal one, of
course. Just for example, we might want to add to our sinusoidal model some underlying
variation, e.g. a simple constant term or a linear or quadratic trend, as it is done in
the generalized least-square periodogram described in \citep{Baluev08a,ZechKur09}. We omit
the detailed formulae for these cases, since they would be relatively bulky for a
presentation here, and the reader can now easily derive them himself.

\section{Practical applicability of the method}
\label{sec_applicability}
Considering a statistical test from the view point of its practical applicability, we
usually ask two main questions,
\begin{enumerate}
\item \emph{Concerning the behaviour of the test under the null hypothesis:} How precisely
we can estimate the false alarm probability, associated with an observed value of the test
statistic?
\item \emph{Concerning the behaviour of the test under the alternative:} Given an accurate
$\FAP$ estimation, how sensitive is our test to practically expected deviations from the
null hypothesis?
\end{enumerate}
Both issues can be addressed by means of numerical simulations, which is the goal of this
section.

\subsection{Test reliability}
For the first issue, we need to specify some model condition satisfying the null
hypothesis, then run a Monte Carlo simulation procedure (generating artifical random
measurement errors), counting how frequently we meet the false alarms (the events when our
test wrongly rejects the null hypothesis), and then to compare the simulated significance
value with the expected one. In our case, we need first to construct some test signal that
produces two alternative observed periods, for which $\KLIC_{12}=0$. It is not so easy as
it may seem, because simultaneously the rival periods should not be indistinguishable in
principle. For example, for strictly evenly spaced observations, $t_i=i\Delta t$, each
periodicity can be equally treated as having the original frequency $\omega$ or any alias
frequency $\omega+2\pi k/\Delta t$ for any integer $k$. All these alternative
interpretations are fully equivalent and observationally indistinguishable under any
circumstances, since they generate exactly the same sequence of values at $t_i$. The Vuong
statistic is not defined for this case at all, because all $l_i$ in~(\ref{li}) are
identically zero. This is not the situation that we want to test, since in practice we
need to compare only potentially distinguishable alternatives. As a more realistic test
model, we can consider a non-degenerate aliasing (the data are gapped in time rather than
strictly evenly spaced), an $\omega_0$-periodic signal, and its two primary aliases
$\omega_{1,2}= \omega_0\pm \omega_g$ (with $\omega_g$ being the data gapping frequency).
If we neglect the main frequency $\omega$, these peer aliases provide practically equal
interpretation of the observations.\footnote{The cases when the true signal period is
never taken into account are not really artifical. For example, the true orbital period of
$0.7$~days for the extrasolar planet 55~Cnc~\emph{e} was hidden for a very long time
behind an alias of $2.8$~days \citep{Dawson10}. This happened only because the researchers
did not plot the periodograms beyond the $1$~day period bound, until recent time.}

Given this test model, we can generate a Monte Carlo sequence of mock time series by
adding simulated random errors to the probe sinusoid and generating random timings
according to the specified gapping pattern. In our case, $N$ timings $t_i$ were
distributed uniformly within $n=10$ periodically gapped intervals $[k P_g, (k+f) P_g]$,
with $f\in [0,1]$ being a filling parameter (fixed during each simulation series). For
each such simulated data set, we evaluate the Vuong statistic $\mathcal V$ comparing the
two primary aliases near $\omega_0\pm\omega_g$. After that, we compare the resulting
simulated distribution of $\mathcal V$ with the expected standard normal distribution.

\begin{figure*}
\includegraphics[height=0.37\textwidth]{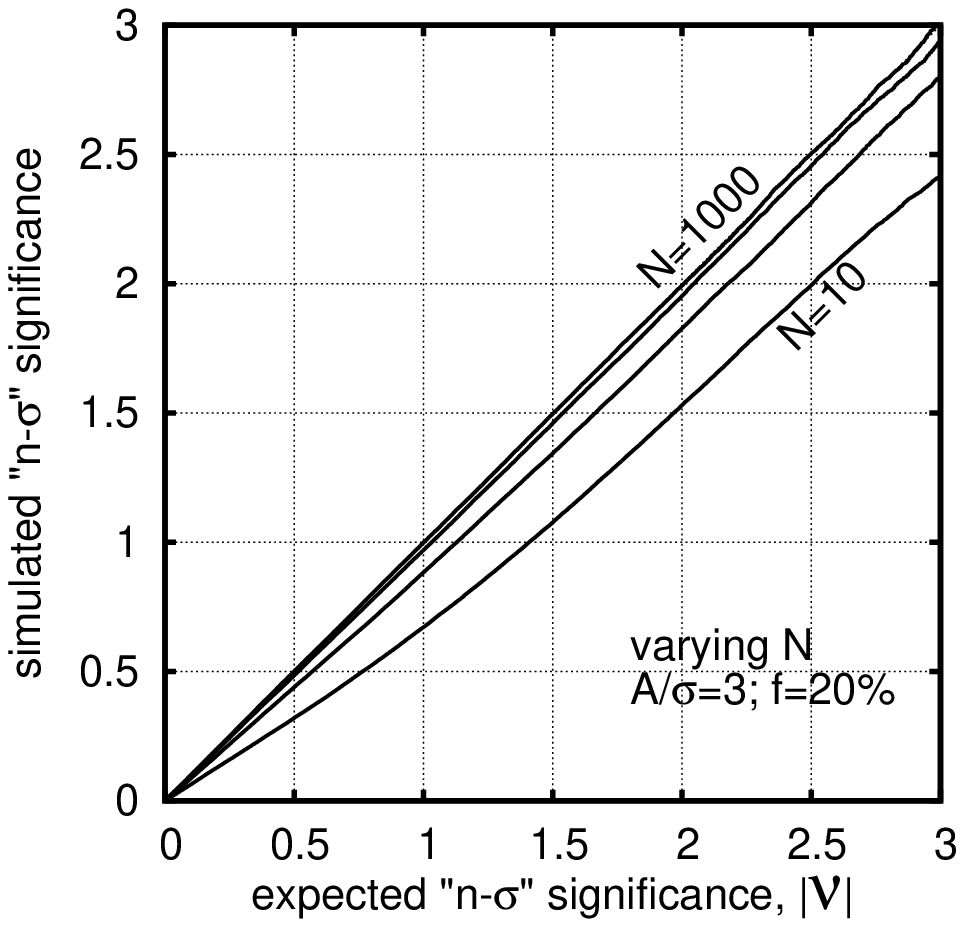}
\includegraphics[height=0.37\textwidth]{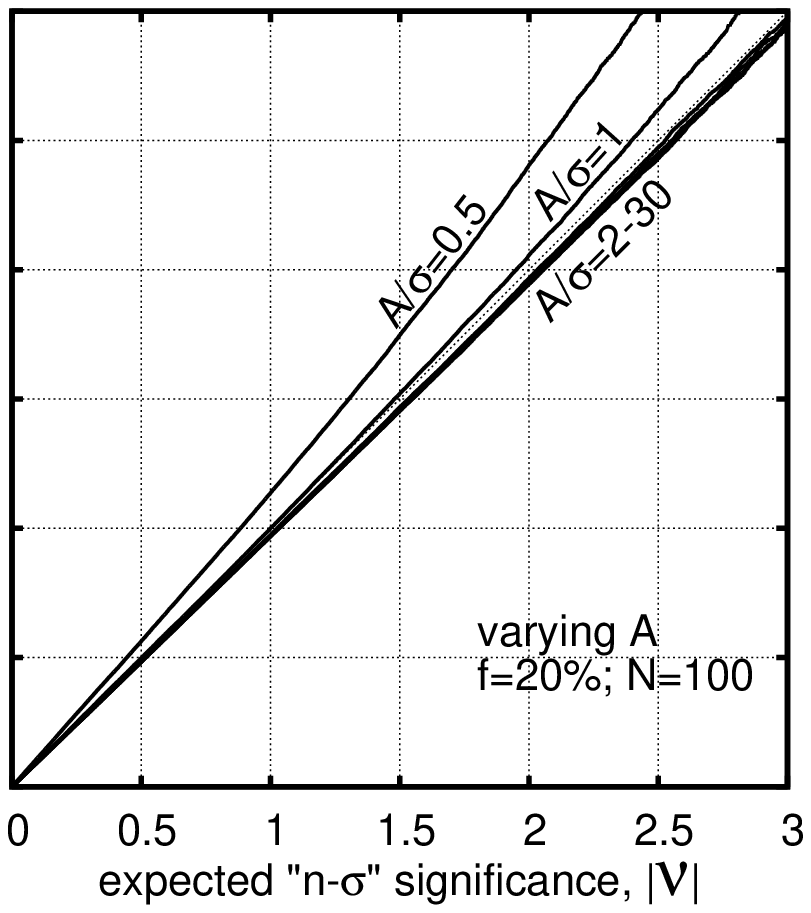}
\includegraphics[height=0.37\textwidth]{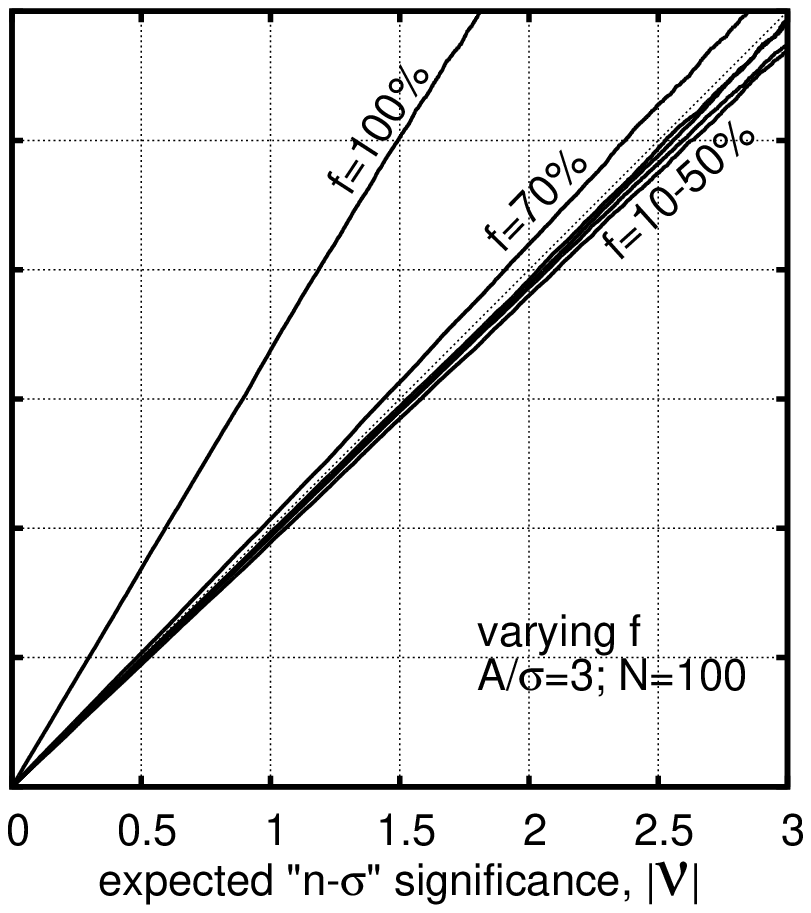}
\caption{Predicted vs. simulated significance levels for the Vuong test, depending on
various parameters. Abscissas show the absolute value of the Vuong statistic, while the
ordinates show the actual (simulated) significance level in the ``$n$-$\sigma$'' notation.
If the Vuong statistic was perfectly normal then the simulated significance level would be
$S(|\mathcal V|)=|\mathcal V|$, and the corresponding graph would strictly follow the
diagonal line. See text for further details.}
\label{fig_null}
\end{figure*}

The results of the simulations are shown in Fig.~\ref{fig_null}. In these diagrams, the
axes show expected and simulated significance levels in the so-called $n$-$\sigma$
notation, which is convenient when dealing with normal or almost normal probabilities. For
the expected significance (in the abscissas) this is just equal to an observed value of
$|\mathcal V|$. Each value in the ordinates, $S$, represents such critical value for a
hypothetical standard normal varaible $x$ that the probability for $|x|$ to exceed $S$ is
equal to the actual simulated probability for $|\mathcal V|$ to exceed the value in the
abscissa. Larger $S$ implies higher significance level and smaller $\FAP$.\footnote{We
remind that the frequently used one-, two-, and three-sigma levels correspond to
$\FAP=31.7\%$, $4.6\%$, and $0.27\%$, and in general $\FAP=2\Phi(S)$.} If $\mathcal V$
indeed follows the expected standard normal distribution then $S(|\mathcal V|) = |\mathcal
V|$, and its graph should strictly follow the main diagonal. If, for instance, $\mathcal
V$ has some non-unit variance $\sigma_{\mathcal V}^2$ and still normal then $S(|\mathcal
V|) = |\mathcal V|/\sigma_{\mathcal V}$.

In Fig.~\ref{fig_null} we show the graphs of $S(|\mathcal V|)$ varying different
parameters of the test model. Overall, the agreement between simulated and expected
significances is good, except for certain rather boundary cases. The bad cases include:
small number of the observations $N\lesssim 30$, very small signal amplitude $A\lesssim
\sigma$, or too weak aliasing $f\gtrsim 70\%$. The two latter bad cases correspond to a
small signal/noise ratio of the aliases involved, so that we even cannot be sure that at
least one of them is actually significant. For example, $f=100\%$ corresponds to no
aliasing at all, when the two rival aliases in our test model are destroyed. Then the test
attempts to compare noisy peaks which appeared in these positions occasionally. Such cases
are not practical, since in practice the significance of at least one of the alternatives
is already established before we ask which one is true.

Summarizing these results, we can derive the following empiric condition of the Vuong test
applicability:
\begin{equation}
A'/\sigma \gtrsim \sqrt{N_0/N},
\label{crit}
\end{equation}
where $A'$ is the best fitting amplitude of the rival periodicities, and $N_0$ is a
constant. The amplitude $A'$ is smaller than the amplitude $A$ of the original generating
signal, since $A'$ refers to the alias periodicities, while the main period is neglected.
For our gapped time series we can readily obtain an analytic approximation for $A'$ using
the classical period analysis formulae \citep[e.g.][]{Vit-nun}:
\begin{equation}
A' \simeq A \frac{\sin \pi f}{\pi f}.
\end{equation}
In practice, it is more convenient to deal with the corresponding periodogram peak value
rather than with the best fitting signal amplitude. We can rewrite~(\ref{crit}) in a very
simle form
\begin{equation}
z' \simeq \frac{{A'}^2 N}{4\sigma^2} \gtrsim z_0=\frac{N_0}{4}.
\label{crit2}
\end{equation}
When applying~(\ref{crit2}) in practice, we should simply check whether the rival
periodogram peaks that we want to compare are both above the $z_0$ level. For out test
problem, we find $N_0\simeq 50$ and $z_0\simeq 10$, although we must note that these
thresholds may increase when e.g. we compare highly non-linear models.

\begin{figure}
\includegraphics[width=84mm]{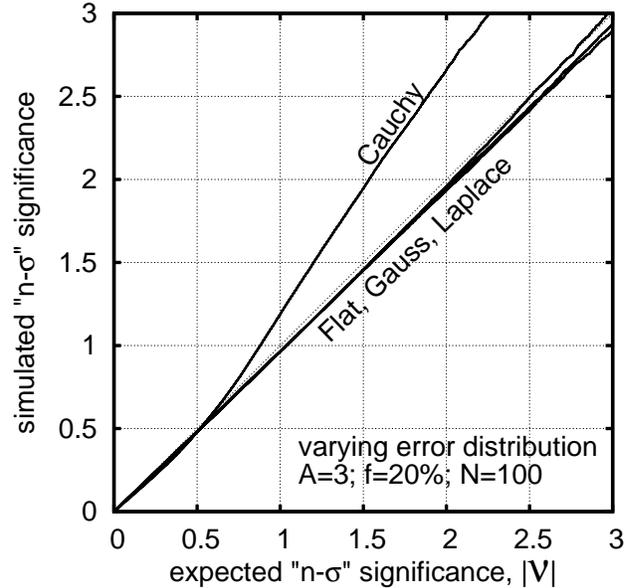}
\caption{Predicted vs. simulated significance levels for the Vuong test, depending on the
true error distribution. The notes from Fig.~\ref{fig_null} apply here. Always assuming
the Gaussian model~(\ref{gauss}) in the Vuong test, we perform simulations using several
non-Gaussian error models: flat p.d.f. in the range $[-1,1]$, Laplace p.d.f. $\propto
e^{-|x|}$, and Cauchy p.d.f. $\propto 1/(1+x^2)$. Only the Cauchy distribution causes
significant effect on the statistic $\mathcal V$.}
\label{fig_null_distr}
\end{figure}

So far we assumed that the measurement errors distribution is always Gaussian. What if in
true this distribution is different from the one used in the Vuong test? Theoretically,
the behaviour of the Vuong test should remain basically the same (in the most cases). This
is verified by the simulations presented in Fig.~\ref{fig_null_distr}. We can see that
most error distribution that we tried out did not introduce any significant changes
indeed. Only the Cauchy distribution produced some large systematic effect. This is not
very surprising, because various pecularities of the asymptotic behaviour are typical for
the Cauchy distribution, due to its heavy tails.

\begin{figure}
\includegraphics[width=84mm]{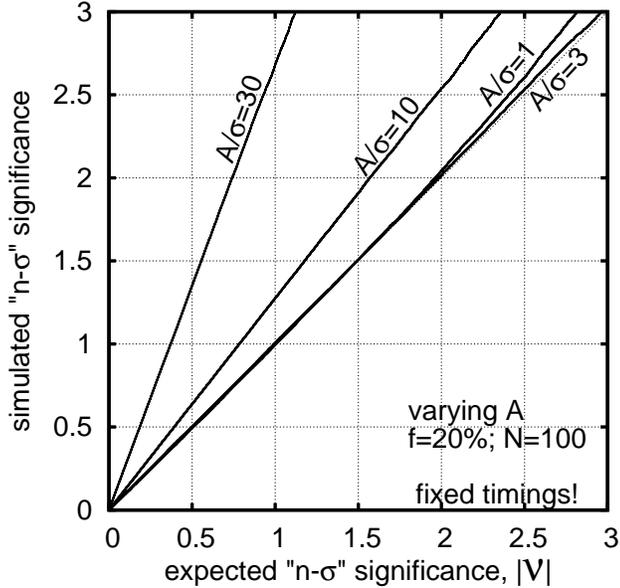}
\caption{Predicted vs. simulated significance levels for the Vuong test, assuming fixed
timings $t_i$. The notes from Fig.~\ref{fig_null} apply here. Here we vary only the signal
amplitude.}
\label{fig_null_A_}
\end{figure}

It is also interesting to investigate the case of fixed $t_i$ (Fig.~\ref{fig_null_A_}). In
this case, the distribution of the Vuong statistic may become significantly more
concentrated, in comparison with the random $t_i$ case. We find that the main critical
parameter in this case is the signal amplitude. For small $A$, the distribution of the
Vuong statistic does not depend much on whether we assume fixed or random timings. For
large $A$, the scatter of the Vuong statistic shrinks, and the simulated significance
level for the fixed $t_i$ becomes larger than for the random $t_i$. This probably occures
because large amplitude scales up the sensitivity of $l_i$ in~(\ref{li}) to the small
fluctuations in $t_i$. It is practically important that the standard normal distribution
still allows us to assess a \emph{lower} bound on the significance of $|\mathcal V|$. This
means that the number of false alarms still remains limited according to our request,
although it may appear smaller. In addition, the cases of large amplitude are not very
important, since in practice a period ambiguity is rare when the signal is so strong. We
would like to emphasize that in practice the timings $t_i$ indeed usually have random
nature that should not be neglected.

\subsection{Test efficiency}
Now let us consider the second main issue~-- the sensitivity of the Vuong statistic to the
cases when one rival model is indeed better than the another one. In this case, we may
consider the same sinusoidal test signal and simply compare the main period with one of
the aliases, e.g. the primary one having frequency $\omega_0+\omega_g$. The average value
of $\mathcal V$ should be now biased to positive values, since the first (correct) model
should look systematically more likely. For each simulation trial, we may have three types
of outcomes: a success ($|\mathcal V|$ exceeds some positive critical value $\mathcal V_*$
and $\mathcal V$ is positive too), a failure ($|\mathcal V|>\mathcal V_*$ but $\mathcal
V<0$), and a neutral (inconclusive) outcome ($|\mathcal V|<\mathcal V_*$). We denote the
respective probabilities as $P_s$, $P_f$, and $P_i$. All of them are functions of the
critical level $\mathcal V_*$, which can be tied to the rejection significance level
$S(\mathcal V_*)\approx \mathcal V_*$ or to the corresponding false alarm probability.
Speaking in terms of Section~\ref{sec_formulation}, the quantities $P_f$, $P_i$, and
$\FAP$ represent the probabilities to make the first, second, or third kind mistake. In
addition, we can now define a few other indicators of the test performance: the
failures/successes probabilities ratio, $\FS$, the failures/inconclusives ratio, $\FI$,
and the successes/inconclusives ratio, $\IS$. Obviously,
\begin{eqnarray}
1/P_s&=&1+\FS+\IS, \nonumber\\
1/P_f&=&1/\FS+1/\FI+1, \nonumber\\
1/P_i&=&1+1/\IS+\FI, \nonumber\\
\FI &=& \FS/\IS.
\end{eqnarray}
Only two variables involved in these equalities can be treated as independent. We choose
to use $\FS$ and $\IS$ as these basic characteristics. These two quantities characterise,
how frequent are mistakes of the first and second kind, in comparison with successful
classifications (given fixed probability of the third kind mistakes, $\FAP$). The quantity
$\FS$ can be also interpreted as a measure of how much our test exceeds a simple drop of a
coin (which has $\FS=1$). The quantity $\IS$ measures the test conservativeness. A perfect
test should possess small values of $\FS$ and $\IS$.

Together with the requested significance level, we now have three independent
characteristics $\FS$, $\IS$, and $S$, which are all functions of a single control
parameter~-- the critical value $\mathcal V_*$. Therefore, to fully characterise the test
performance, we should investigate the corresponding parametric curve in the
three-dimensional space $(\FS,\IS,S)$. This is a bit more complicated than, for example,
in the usual signal detection task, when we have only two independent variables~--- the
false alarm probability and the probability of wrong non-detection. We therefore have no
other option than to deal with some two-dimensoinal projections. Since the bahaviour of
$S(\mathcal V_*)\approx \mathcal V_*$ has been already investigated in details, we now
look at the pair $(\FS,\IS)$.

\begin{figure*}
\includegraphics[height=0.37\textwidth]{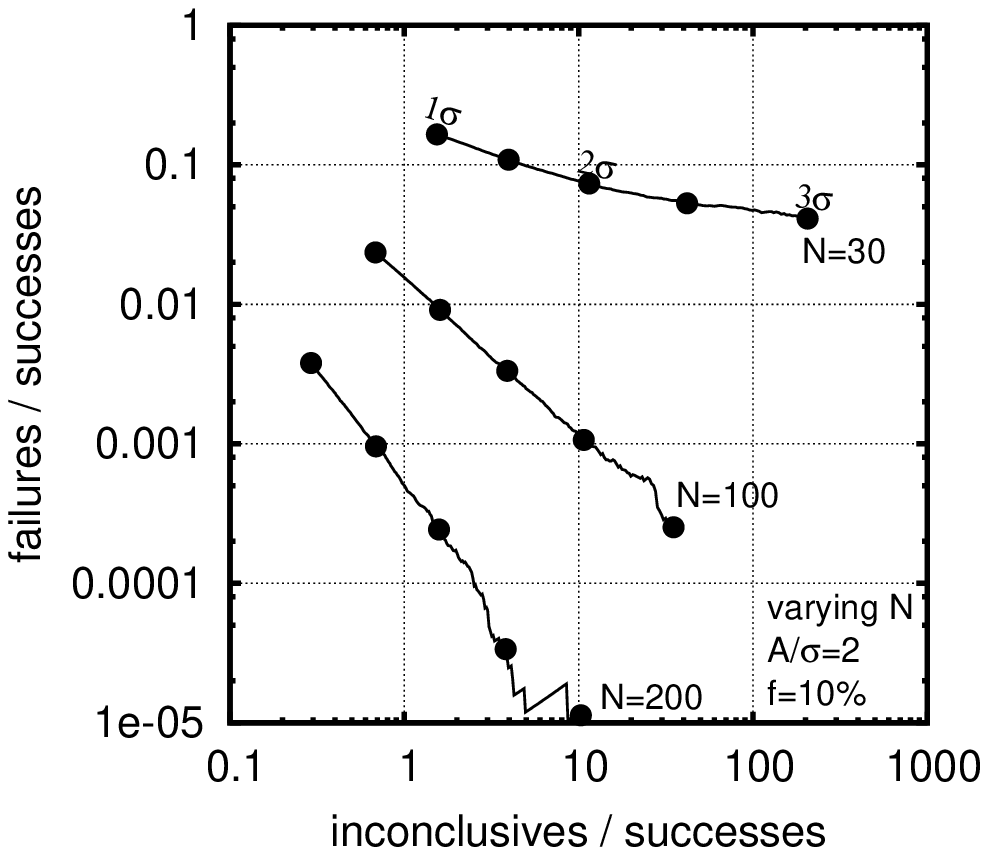}
\includegraphics[height=0.37\textwidth]{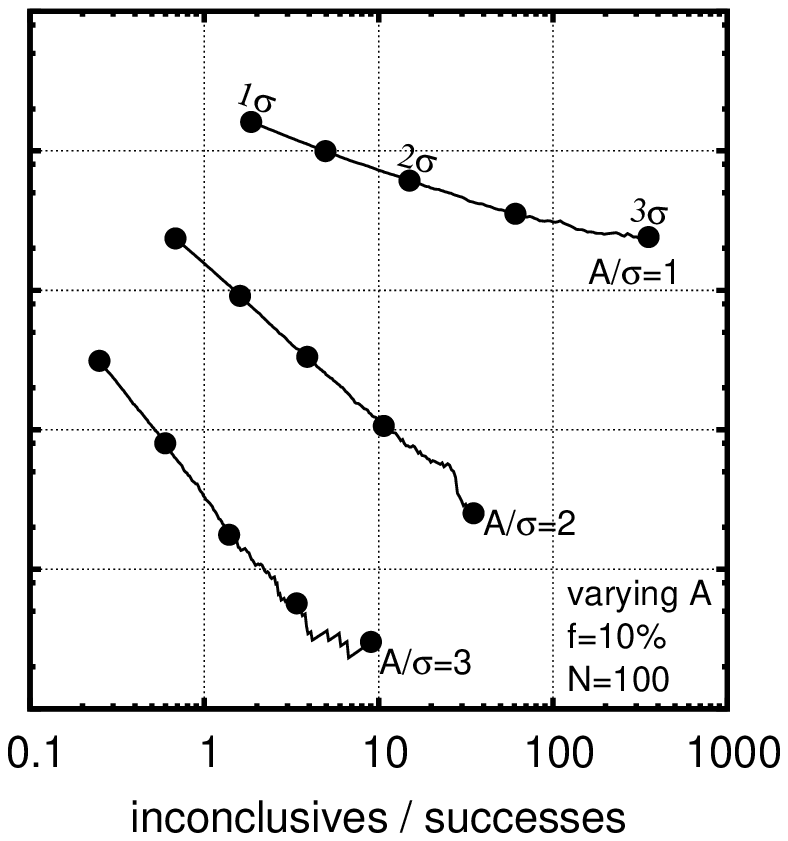}
\includegraphics[height=0.37\textwidth]{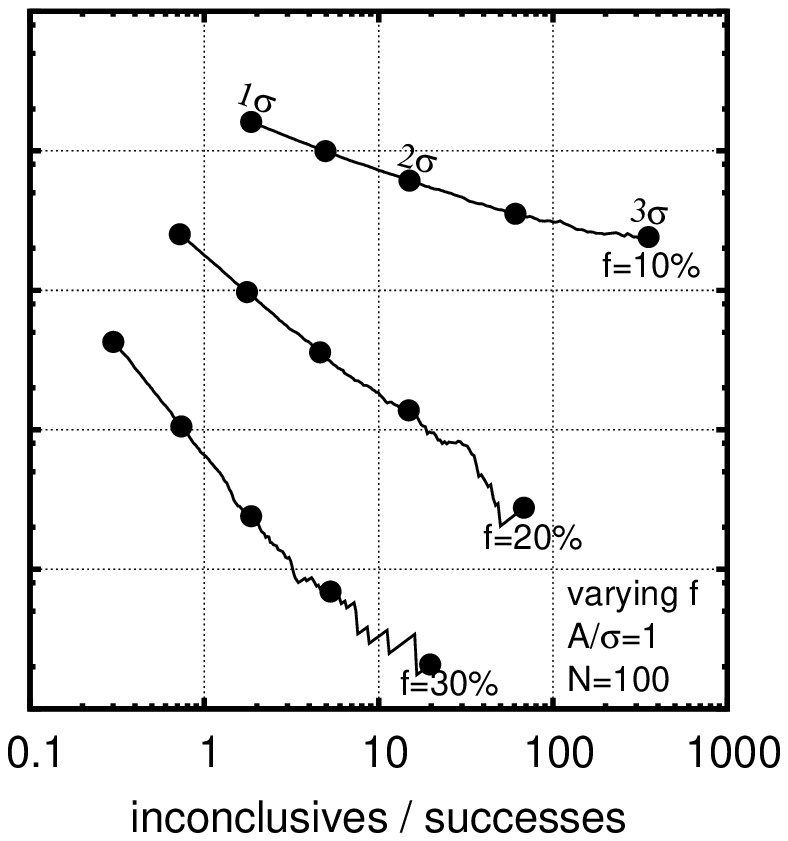}
\caption{Simulated performance of the Vuong test, depending on various parameters. Here we
inspect the behaviour of the Vuong statistic comparing the primary (``true'') peak near
the main frequency $\omega_0$ and its alias near $\omega_0+\omega_g$ (the exact values
were adjusted to achieve a local maximum of the periodogram). In each curve we also put
five points marking the positions of $\mathcal V_*=1,1.5,2,2.5,3$, which approximately
correspond to the same significance levels ($S$), i.e. from one-sigma to three-sigma. The
``failures'' refer to the cases when the Vuong test suggested to select the alias peak.
The number of Monte Carlo simulations was $10^6$. See text for further details.}
\label{fig_perf}
\end{figure*}

In the Fig.~\ref{fig_perf} we plot several graphs of $(\IS(\mathcal V_*),\FS(\mathcal
V_*))$ as parametric curves, setting different values for the signal and time series
parameters. On each curve we also mark a few points corresponding to the values of
$\mathcal V_*$ from $1$ to $3$. As we could expect, increasing the signal amplitude $A$,
or the number of observations $N$, or the time series filling factor $f$ decreases both
the probability of a misclassification and of an inconclusive result. We may note that
when varying different control parameters we usually obtain very similar curves. This
indicates that there should be a single quantity (a combination of $N$, $A$, and $f$) that
defines the test performance for our task. On the basis of the presented simulations, we
can empirically construct this critical quantity as
\begin{equation}
G \simeq \frac{A}{\sigma} f \sqrt N.
\label{thr}
\end{equation}
Three curves in each panel of the Fig.~\ref{fig_perf} correspond to $G\approx 1$,
$G\approx 2$, and $G\approx 3$. The formula~(\ref{thr}) can be also justified
theoretically. Indeed, since the Vuong statistic is based on the quantity $L$, which is
expressible as the difference~(\ref{sumli}) between two rival periodogram values, the
latter difference should represent a critical parameter characterising the test
performance. For the primary peak and its alias, this difference looks like
\begin{eqnarray}
z_1-z_2 \simeq N \frac{A^2 - {A'}^2}{4\sigma^2} \simeq
\frac{A^2 N}{4\sigma^2} \left[1 - \left(\frac{\sin \pi f}{\pi f}\right)^2 \right] \simeq \nonumber\\
\simeq \frac{A^2 \pi^2 f^2 N}{12\sigma^2} = \frac{\pi^2}{12} G^2.
\end{eqnarray}
This is not a rigorous proof of~(\ref{thr}), however.

We can see that the Voung test can manage the misclassification errors very well, but it
favours a relatively large number of inconclusive results. This is not very surprising,
since this test is originally designed to be conservative in drawing definite conclusions.
This is the main reason why it suppresses the misclassifications so efficiently. From the
presented simulation results, we may also note that in practice it does not make much
sense to request very high significance level from the Vuong test. Requesting large value
of $S(\mathcal V_*)$ can further suppress the third-type mistakes, but by the cost of a
dramatic increase in the fraction of inconclusive results, which can even become
dominating. Simultaneously, the relative fraction of the first-type mistakes
(misclassifications) either is not decreased very much or is already sufficiently small
for moderate $S$. Therefore, we believe that in practice it is enough to set the required
significance level about $1.5-2$ sigma.

\begin{figure*}
\includegraphics[height=0.37\textwidth]{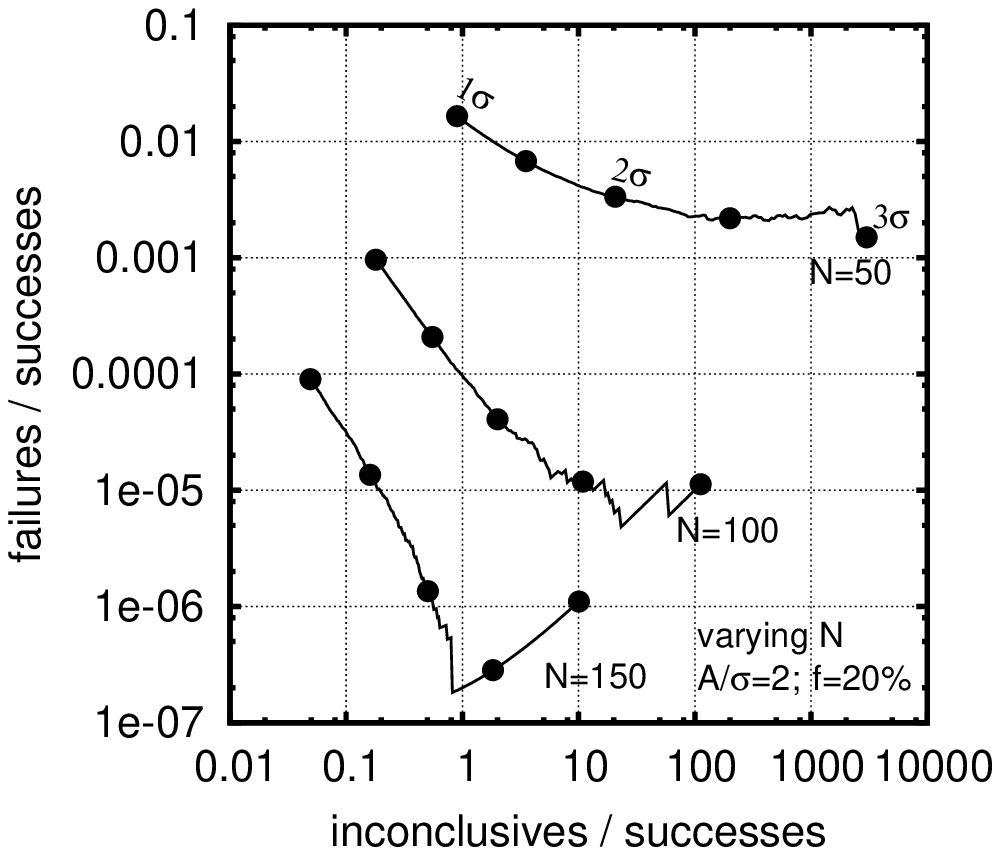}
\includegraphics[height=0.37\textwidth]{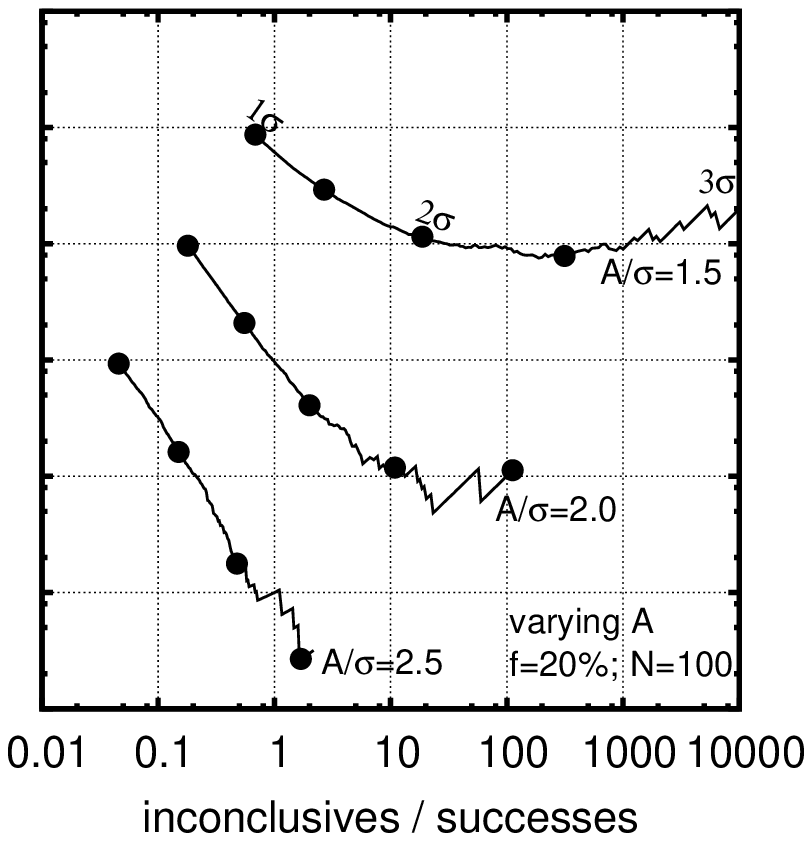}
\includegraphics[height=0.37\textwidth]{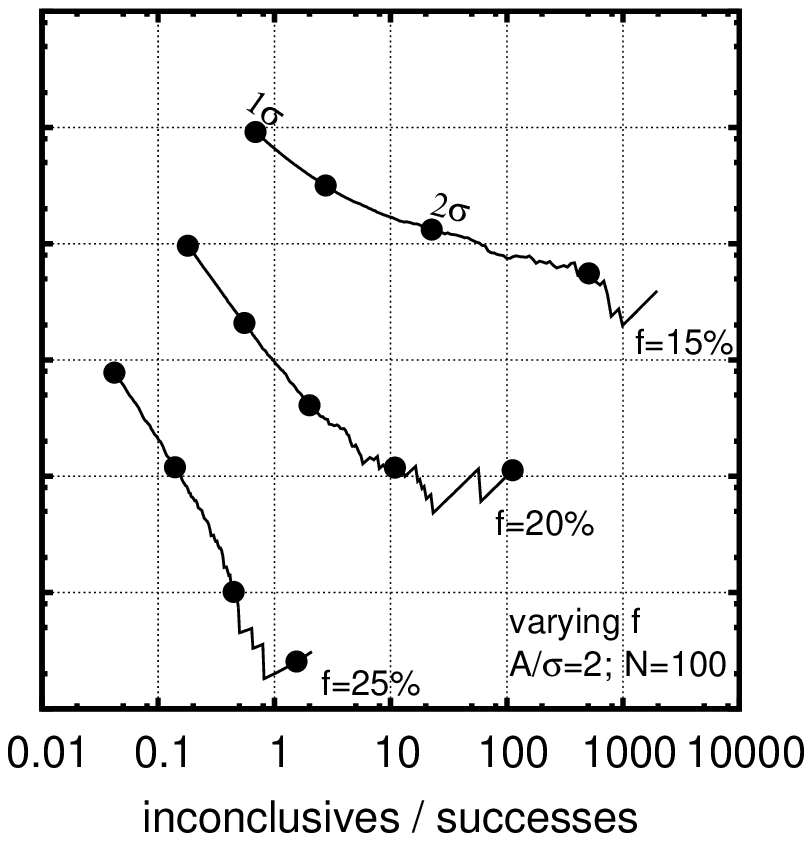}
\caption{Simulated performance of the Vuong test, depending on various parameters. Similar
to Fig.~\ref{fig_perf}, but instead of picking two given periodogram peaks (the main one
and a given alias) we now inspect two \emph{largest} peaks among \emph{many} period
candidates near $\omega_0+k\omega_g$. The parts of the curves corresponding to $\mathcal
V_*$ values between $2.5$ and $3$ are very unreliable here, regardless a larger number of
Monte Carlo simulations ($10^7$). These curves appeared apparently the same for two
different definitions of a ``failure'': (i) the cases when the Vuong test recommends to
select a peak, while the second candidate is actually better, and (ii) the cases when the
Vuong test recommends to select any alias peak (not the true one). In the second
interpretation, selecting a $k^{\rm th}$-order alias was counted as $k$ failures at once.
The plotted curves are for the first interpretation. See text for further details.}
\label{fig_perf_m}
\end{figure*}

In practice we often deal with more complicated situations, however. In addition to the
true period, we may have also \emph{multiple} aliases, and there is no guarantee that the
true period is indeed associated to one of the two largest periodogram peaks. Both
selected highest peaks may occure aliases. In that case, we need to answer a principal
question: what event we should treat as a failure of the test? Apparently, a failure
should occur each time when the test suggests to choose an alias period. However, testing
for multiple alternatives is not the responsibility of the Vuong test. The only thing that
we required from this test, by its design, is to choose a \emph{better} solution among the
\emph{two} ones provided on input. From this view point, the failures only occur when the
test suggests to choose the peak which is more distant from the true period than the
alternative candidate. The cases when the test selects an alias peak, while the second
candidate is a larger-order alias, should be treated as successes (even if the selected
peak is an alias too).

We performed simulations for both interpretations. To increase the contrast, we also added
some penalty in the first interpretation, counting each selected $k$-order alias near
$\omega = \omega_0\pm k\omega_g$ as $k$ failures at once. The results appeared apparently
the same, however. This indicates that the cases, when both the largest periodogram peaks
are only aliases, are rare and usually trigger an inconclusive decision of the Vuong test.
The results of the simulations are shown in Fig.~\ref{fig_perf_m}. In fact, this figure
differs from Fig.~\ref{fig_perf} mainly because it is now allowed to choose any of the two
first-order aliases near $\omega_0\pm \omega_g$, instead a single alias alternative
$\omega_0+\omega_g$. A high-order alias may be selected too, but, as we have just
discussed, the probability of such an event appears negligibly small. Comparing
Fig.~\ref{fig_perf_m} with Fig.~\ref{fig_perf}, we note two main differences. First, the
fraction of inconclusive results is increased, as well as its sensitivity of the requested
value of $\mathcal V_*$. Second, the threshold value of $G$, necessary to suppress these
inconclusive decisions, grew roughly by half, and the sensitivity of the simulated curves
to the value of $G$ is also increased. To have, say, only half of inconclusive decisions,
given $1.5$-sigma or $2$-sigma confidence level ($S$), the value of $G$ should be about
$4.5$ or $3.5$.

\section{Applications}
\label{sec_exo}
\subsection{Exoplanetary system orbiting 55~Cnc}
For many years it was believed that the planet 55~Cnc~\emph{e}, the innermost planet in
the 55~Cnc system, posesses an orbital period of $P_e\approx 2.8$~day, until
\citet{Dawson10} showed that this apparent periodicity is actually a diurnal alias of the
true one with $P_e\approx 0.7$~day. This new period value allows for a significantly
better fit of the available data. This true period value could hide for so long time only
because the researchers did not consider any potential periods smaller that $1$~day.
Here we would like to rigorously compare these alternative period values utilising our new
method. We use Lick and Keck radial velocity data from \citep{Fisher08} to obtain two
alternative five-planet fits with $P_e\approx 0.7$~day or $P_e\approx 2.8$~day. With the
use of multi-Keplerian (unperturbed) model of the radial velocity we obtain $\mathcal
V\approx 5.6$ for these alternatives. Using $N$-body Newtonian radial velocity model, as
described in \citep{Baluev11}, we have $\mathcal V\approx 5.5$ (assuming the system is
seen edge-on). No doubts, the period of $0.7$~day is indeed the correct one.

We may also carry out another, somewhat unusual, model comparison. Estimated planetary
orbital parameters and masses, as well as the fit quality, are a bit different for
Keplerian and Newtonian models of radial velocity. So we might ask: are these differences
statistically significant? In other words, can we say that mutual planetary perturbations
are already detected in the RV data? To answer this question, we just need to find two
orbital fits, one using Keplerian and another using Newtonian model, and then to calculate
the Vuong statistic for these models. We obtain $\mathcal V$ of only $0.2$. Allowing for
the common orbital inclination to the sky plane to float during the fitting, we get
$\mathcal V\approx 0.4$ (with the best fitting inclination of $16^\circ$). So small values
of $\mathcal V$ say that possible gravitational perturbations in this system are not yet
detectable from RV curve. Therefore, any attempt to constrain orbital inclinations in this
system form the RV data (on the basis of potential interplanetary perturbations) would be
probably meaningless, possibly except for putting some very mild and thus not too much
useful limits.

\subsection{Exoplanetary system orbiting HD75898}
\begin{figure*}
\includegraphics[width=0.85\textwidth]{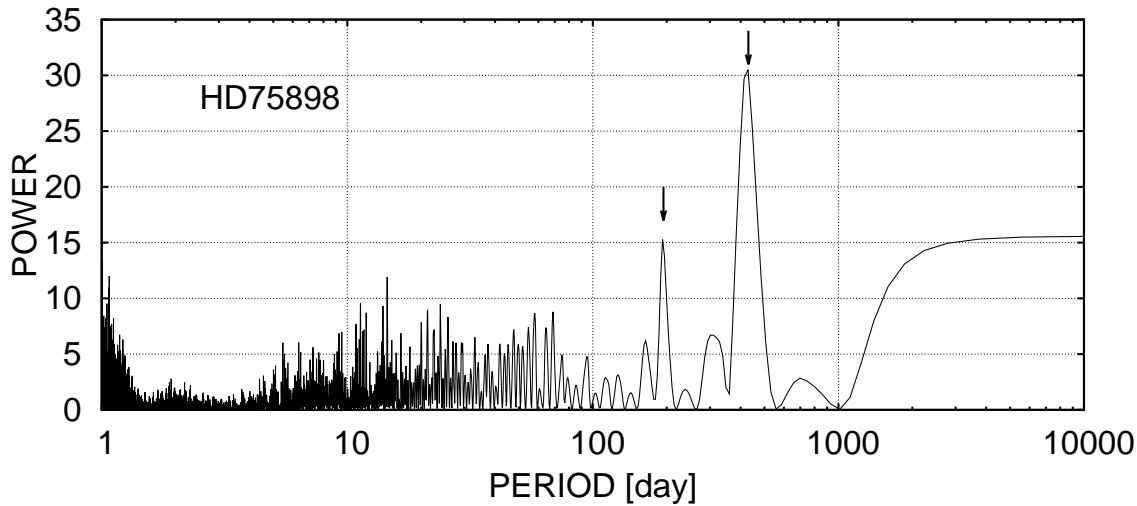}
\caption{Periodogram of HD156668 radial velocity data, showing two high peaks at the
period values of $400$~days and $200$~days. The Vuong test suggests that these peaks are
distinguishable at $3.5\sigma$ level, thus we may definitely accept the peak at $400$~days
as the correct solution for the planet \emph{b}. Here we perfectly agree with
\citet{Dawson10}.}
\label{fig_pow_HD75898}
\end{figure*}

This star was observed by the Keck team \citep{Robinson07}, who reported the discovery of
a Jovian-mass planet having orbital period of about $400$~days. The relevant periodogram
shows actually two comparable peaks at the periods of $400$~days and $200$~days (see
Fig.~\ref{fig_pow_HD75898}). Let us try to verify the results by \citet{Dawson10}, who
investigated this aliasing ambiguity more carefully and confirms that the $200$-day period
should be rejected indeed. We basically agree: $\mathcal V \approx 3.5$ in this case, so
we can safely choose the period of $400$~days, since the difference between the models has
very high significance of more than $3\sigma$.

\subsection{Exoplanetary system orbiting GJ876}
The planetary system orbiting GJ876 is currently believed to host four planets
\citep{Rivera10}. The detailed analysis of all up-to-date radial velocity data for this
system, including recent HARPS \citep{Correia10} and Keck \citep{Rivera10} data, is given
in \citep{Baluev11}, along with the full orbital configuration details. Here we would like
to consider two ambiguities associated to this planetary system.

The first issue is related to the orbital period of the innermost planet GJ876~\emph{d}.
Since the very discovery of this planet it was noted \citep{Vogt05} that (presumably) the
primary period value $P_d\approx 1.938$~day is accompanied by a diurnal alias of
$P_d\approx 2.055$~day. The analysis done by \citet{Dawson10} supports this
classification, as well as our calculations, which yield $\mathcal V\approx 2.0$ for this
alias ambiguity. Therefore, here we are able to rigorously confirm that these periods
indeed are well-distinguishable and the correct period is $1.938$~day.

The second issue that we would like to highlight concerns the determinability of the
planet GJ876 \emph{e} eccentricity, $e_e$. It is noted in \citep{Baluev11} that this
eccentricity, although is bounded by $\sim 0.2$ from the upper side, looks still
ill-determined below this limit. In particular, we have two comparable local minima of the
likelihood function: the first one at $e_e\approx 0$ and another at $e_e\approx 0.15$
(with the corresponding pericenter argument $\omega_e \sim 45^\circ$). Therefore, this
ambiguity looks like a good task suitable for the statistical methods proposed in the
present paper. We find that the Vuong test agrees that these two orbital solutions are
observationally indistinguishable: $\mathcal V$ varies within the range of $0.5-1$,
depending on some minor model details. This result allows us to confirm more rigorously
the conclusion that $e_e$ is indeed ill-determined.

As it is shown in \citep{Baluev11}, both HARPS and Keck data for GJ876 contain significant
autocorrelated component (red noise). We must acknowledge that this red noise could make
the Vuong test not very reliable, because such measuremets are not uncorrelated.
Nevertheless, in practice taking this red noise into account usually \emph{increases}
various statistical uncertainties, so apparently distingushable models could appear
actually equivalent, but hardly vice versa.

We do not believe, however, that this red noise could affect our previous disambiguation
of the planet \emph{d} period. The two alternatives for this period reside very close to
each other, while the effect of the correlated noise is spread over the whole frequency
spectrum. It could distort the balance between some distant periodogram peaks, but not
between these ones.

\subsection{Exoplanetary system orbiting GJ3634}
\begin{figure*}
\includegraphics[width=0.85\textwidth]{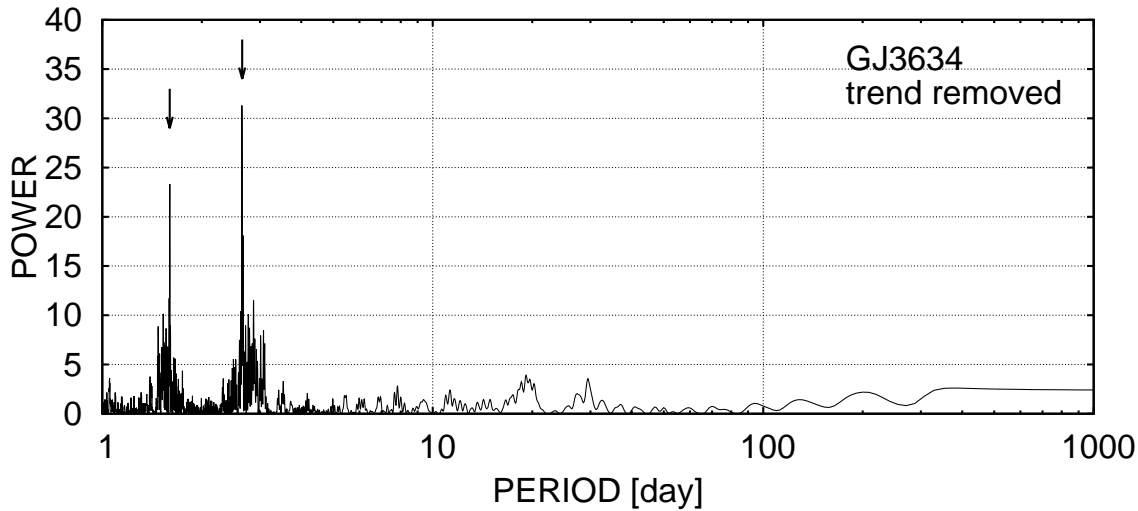}
\caption{Residual periodogram of GJ3634 radial velocities, with long-term quadratic drift
removed. We can see two peaks here. The peak at $2.6$~days is actually double. The Vuong
test suggests to select the higher peak at $2.6$~day in comparison with the peak at
$1.6$~day, but at a moderate significance of $1.8\sigma$. The smaller peak near $2.6$~day
can be rejected at $2.4\sigma$.}
\label{fig_pow_GJ3634}
\end{figure*}

\begin{table}
\caption{Comparing three alias alternatives for GJ3634~\emph{b}: values of
$\mathcal{V}$ obtained for various alias pairs}
\label{tab_GJ3634}
\begin{tabular}{ccccc}
Alias         & $1.60$~d & $2.67$~d \\
\hline
$2.65$~d      & $1.8$    & $2.4$     \\
$1.60$~d      & ---      & $0.8$     \\
\hline
\end{tabular}
\end{table}

According to \citet{Bonfils11}, this star is orbited by a super-Earth planet each
$2.6$~day. As the discoverers note, in addition to the periodic signature of this planet,
the radial velocity data also contain a parabolic long-term drift. The periodogram of the
RV data with this trend removed (that is, included in the base model) is shown in
Fig.~\ref{fig_pow_GJ3634}. This periodogram shows two apparent peaks, with the peak at
approximately $2.6$~days being actually double, consisting of a close pair at the periods
of $2.65$~day (height $32$) and $2.67$~day (height $18$). The discovery team does mention
the period of $1.6$~day, but they just retract it as an alias without any rigorous
justification. They also do not mention that the peak at $2.6$~day is double. The results
of our analysis are given in Table~\ref{tab_GJ3634}, where each cell contains a numerical
value of $\mathcal V$, comparing the aliases marked in the left column and in the top
line. We can see that the peak at $2.67$~day can be safely retracted in favour of its
neighbour at $2.65$~day (significance $2.4\sigma$). The period value of $1.6$~day is also
rather unlikely, albeit now $\mathcal V\approx 1.8$, which is only moderately significant.

\subsection{Exoplanetary system orbiting HD156668}
\begin{figure*}
\includegraphics[width=0.85\textwidth]{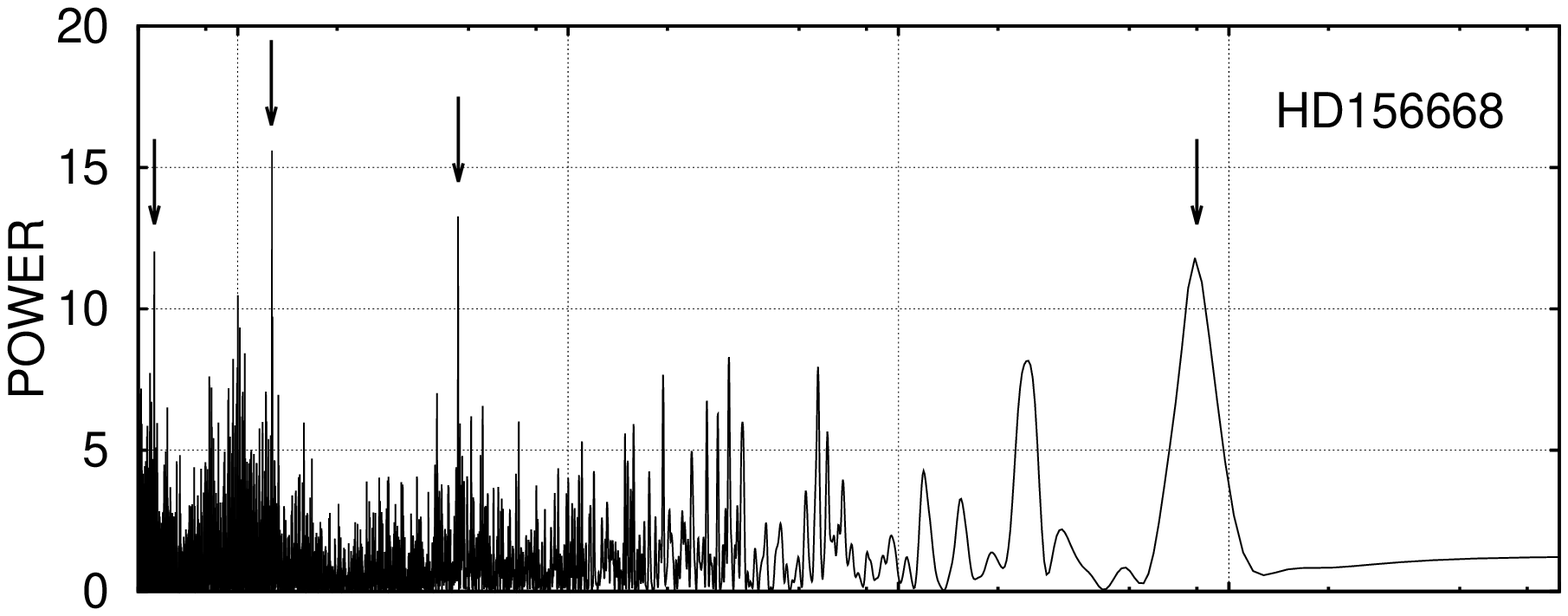}
\includegraphics[width=0.85\textwidth]{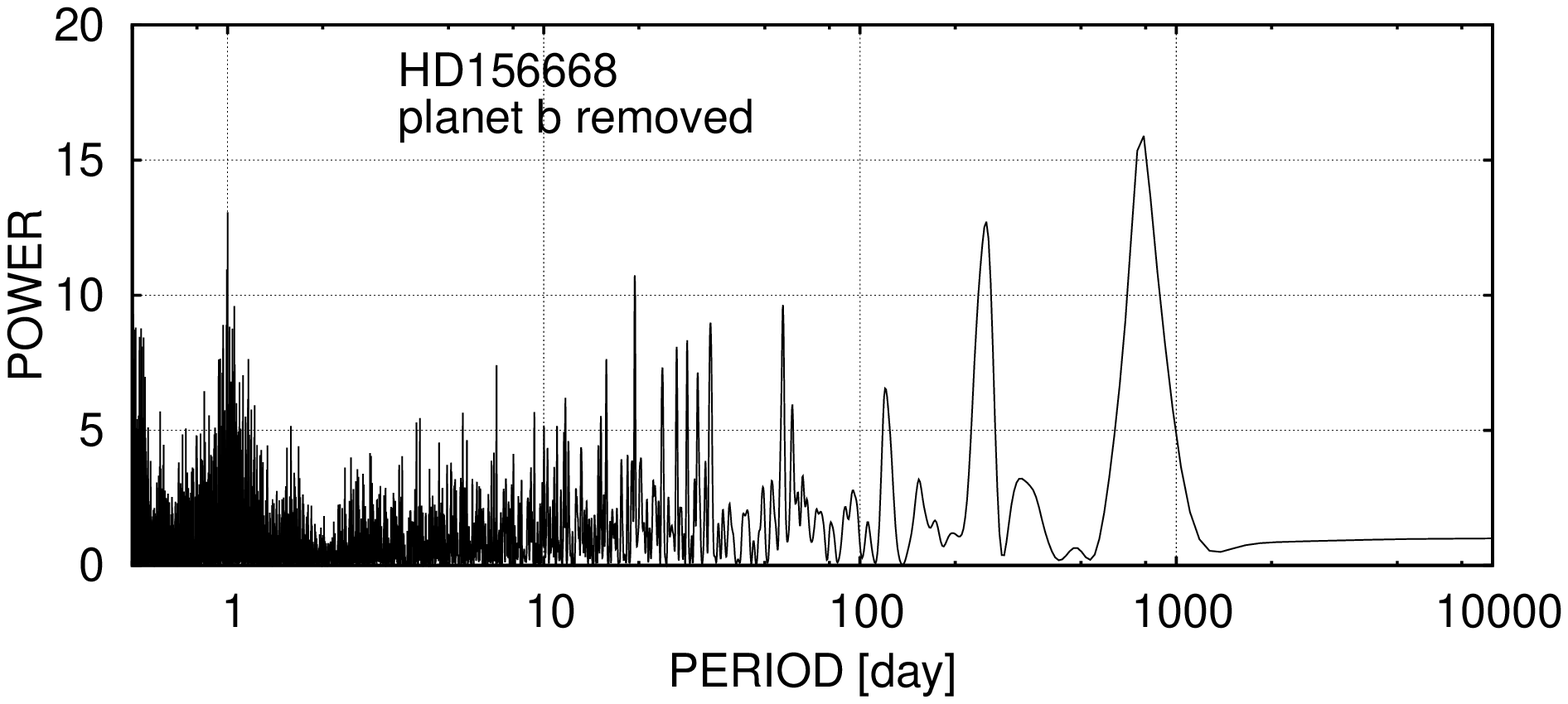}
\caption{Top: periodogram of HD156668 radial velocity data calculated according to
\citep{Baluev08b} and involving only a constant term in the base model. Bottom: similar
residual periodogram with planet \emph{b} sinusoidal signature embedded in the base model
as well (and assuming the period $P_b\approx 1.2$~days). In the second periodogram, the
peak at $800$~days remains intact (and even enforced). It has the same significance as the
$1.2$-day peak in the first periodogram. Here we disagree with \citet{Dawson10}, who
suggested the $1.2$~days peak as a definite solution.}
\label{fig_pow_HD156668}
\end{figure*}

On the basis of the Keck observations, \citet{Howard11} reported that this star is orbited
by a super-Earth having short period of $4.6$~days or, possibly, $1.2$~days. As
\citet{Dawson10} claim, the correct value of the period should be $1.2$~days, and
the peak at $4.6$~day is only its diurnal alias. Let us try to verify these results too.
First, we can see that there are actually more that two possible periods at
stack~(Fig.~\ref{fig_pow_HD156668}, top). For further investigation we select four
periodogram peaks for periods longer than half-day. We fit four corresponding models
assuming circular orbit\footnote{The orbital eccentricity looks ill-determined in this
case, because it often rises to unrealistic values. Also, due to the famous tidal orbit
circularization effect, it is unlikely that an eccentric orbit could reside so close to
the star.}
and calculate the Vuong statistic for each pair of these models. The results are given in
Table~\ref{tab_HD156668}. Investigating this table, we can see that the Vuong statistic is
always small ($<1$). Thus our results do not support the conclusion made by
\cite{Dawson10} concerning this planet.

\begin{table}
\caption{Comparing four alias alternatives for HD156668~\emph{b}: values of
$\mathcal{V}$ obtained for various alias pairs}
\label{tab_HD156668}
\begin{tabular}{ccccc}
Alias         & $4.6$~d & $0.56$~d & $800$~d \\
\hline
$1.2$~d       & $0.9$   & $0.8$    & $0.5$   \\
$4.6$~d       & ---     & $0.3$    & $0.2$   \\
$0.56$~d      &         & ---      & $0.0$   \\
\hline
\end{tabular}
\end{table}

We prefer to remain cautious also because it seems that some extra variations may be
present in the radial velocity of this star. The residual periodogram with, say, $1.2$-day
oscillation subtracted off, demonstrates that the peak at $800$~days does
\emph{not} disappear (Fig.~\ref{fig_pow_HD156668}, bottom). Its height (and thus
significance) appears the same as for the first period that we extracted. Indeed, using
the methods described in \citep{Baluev08a,Baluev08b} we find that this peak infers the
false alarm probability of only $~0.1\%$. It does not matter if we try to extract first
the periodicity of $4.6$~day or $0.56$~day instead of $1.2$~day, the $800$-day peak
remains here. On contrary, extracting this $800$-day peak at the first place does not
eliminate the three remaining peaks at short periods. Speaking shortly, three short
periods are mutual aliases of each other, while $800$-day period is a standalone entity.

This long-period variation might represent a hint of an extra planet, or maybe of some
other phenomenon, like the red noise effect in the RV data for GJ876 \citep{Baluev11}.
Indeed, similarly to the GJ876 case, we can see in this periodogram a broad band of
low-frequency peaks without any standalone clearly dominating peak (either $800$ days or
any other). Just like any usual harmonic periodicity, this low-frequency band produces a
diurnal alias band around the period of $1$~day. This diurnal band is actually clearly
seen in the periodogram, and it might easily enforce the $1.2$-day peak, relatively to the
$4.6$-day one. This argumentation, along with a small value of the Vuong statistic, forces
us to disobey the recommendation by \citet{Dawson10} to leave only $1.2$~day period as the
correct solution. We do not find a rigorous statistical basis for this in the current
data.

\section{Conclusions and discussion}
We have introduced the Vuong test, which is an asymptotically normal statistical test for
verifying the observational equivalence of alternative models. This method compare the
alternative models in the sense of certain statistical measure of their divergence. The
Vuong statistic uses the Kullback-Leibler Information Criterion to establish whether the
models are equivalent or not. Importantly, this test does not require that one of the
rival models should be correct. Instead, both alternatives may occur formally wrong
(misspecified case). Such models are still correctly ordered in terms of the adopted
divergence measure. We classify this as a high practical advantage, since in practice no
model is precisely correct: any data always contain some hidden residual variations and
the true error distribution may deviate from the normal function. We should however admit
that any ordering of the alternatives is not very useful if all available alternatives are
too far from the truth. Another important advantage of our test is due to its practical
simplicity and calculation speed. It is fully analytic and do not require any
CPU-consuming simulations in the routine use.

Throughout the paper, we mainly focused our attention on the period ambiguity issue, since
this is one of the most practical task. However, the Vuong test can be also applied to
check the model equivalence in other situations, when the ambiguity is related to any
other parameters (not necessarily the period). The main necessary formulae of the test
remain basically the same.

We have reanalysed the radial velocity data for several extrasolar planetary systems. We
verified that orbital periods of the planets 55~Cnc~\emph{e}, HD75898~\emph{b} and
GJ876~\emph{d} can be resolved unambigiously on the basis of presently available data,
while the period of HD156668~b and the eccentricity of the planet GJ876~\emph{e} are still
ambigious.

A popular rival approach for solving such disambiguation tasks is the Bayesian model
selection \citep[e.g.][]{Gregory07a}. However, the great disadvantage of the Bayesian
methods is their computational complexity requesting for intensive numerical simulations.
The Vuong test is completely free from this disadvantage. Also, when dealing with a task
like distinguishing between different alternatives, we should always have an adequate
impression of which statement is derived from the real observations, and which one follows
from an a priori assumption not tied to the actually observed data. This impression is
difficult to obtain if we use Bayesian statistics, especially when we compare two
solutions from very different regions of the parameter space. Basically, Bayesian methods
mix the subjective prior assumptions with the objective information derived from the real
data, which makes them, in our view, often unsuitable for model intercomparison. We do not
decline the known strengths of the Bayesian methods, but we must mention that debates on
the Bayesianism still do not cease \citep{Efron86,Gelman08}. In practice, it is always
necessary to verify that the results of the Bayesian analysis are stable with respect to
choosing different reasonable prior distributions. If this stability is not ensured, then
there is a risk that such analysis is in fact more based on the prior assumptions rather
than on the real data.

The Vuong test is not Bayesian, but it does not implicitly assume any specific prior
distributions. Although subjective assumptions are never purged out completely, we can
pursue to minimize their influence. The Vuong test attains good resistence to this
undesired influence. We count it as its great advantage.

\section*{Acknowledgments}
This work was supported by the Russian Academy of Sciences research programme
``Non-stationary phenomena in the objects of the Universe''. I am grateful to the
reviewers for their insightful comments on this manuscript.

\bibliographystyle{mn2e}
\bibliography{aliases}

\begin{thebibliography}{}

\bibitem[\protect\citeauthoryear{Baluev}{Baluev}{2008}]{Baluev08a}
Baluev R.~V.,  2008, MNRAS, 385, 1279

\bibitem[\protect\citeauthoryear{Baluev}{Baluev}{2009}]{Baluev08b}
Baluev R.~V.,  2009, MNRAS, 393, 969

\bibitem[\protect\citeauthoryear{Baluev}{Baluev}{2011}]{Baluev11}
Baluev R.~V.,  2011, Celest. Mech. Dyn. Astron., 111, 235

\bibitem[\protect\citeauthoryear{Bonfils, Gillon, Forveille, Delfosse, Deming,
  Demory, Lovis, Mayor, Neves, Perrier, Santos, Seager, Udry, Boisse \&
  Bonnefoy}{Bonfils et~al.}{2011}]{Bonfils11}
Bonfils X.,  Gillon M.,  Forveille T.,  Delfosse X.,  Deming D.,  Demory B.-O.,
   Lovis C.,  Mayor M.,  Neves V.,  Perrier C.,  Santos N.~C.,  Seager S.,
  Udry S.,  Boisse I.,    Bonnefoy M.,  2011, A\&A, 528, A111

\bibitem[\protect\citeauthoryear{Correia, Couetdic, Laskar, Bonfils, Mayor,
  Bertaux, Bouchy, Delfosse, Forveille, Lovis, Pepe, Perrier, Queloz \&
  Udry}{Correia et~al.}{2010}]{Correia10}
Correia A. C.~M.,  Couetdic J.,  Laskar J.,  Bonfils X.,  Mayor M.,  Bertaux
  J.-L.,  Bouchy F.,  Delfosse X.,  Forveille T.,  Lovis C.,  Pepe F.,  Perrier
  C.,  Queloz D.,    Udry S.,  2010, AA, 511, A21

\bibitem[\protect\citeauthoryear{Cox}{Cox}{1962}]{Cox62}
Cox D.~R.,  1962, J. Roy. Stat. Soc. B, 24, 406

\bibitem[\protect\citeauthoryear{Dawson \& Fabrycky}{Dawson \&
  Fabrycky}{2010}]{Dawson10}
Dawson R.~I.,  Fabrycky D.~C.,  2010, ApJ, 722, 937

\bibitem[\protect\citeauthoryear{Efron}{Efron}{1986}]{Efron86}
Efron B.,  1986, American Statistician, 40, 1

\bibitem[\protect\citeauthoryear{Fischer, Marcy, Butler, Vogt, Laughlin, Henry,
  Abouav, Peek, Wright, Johnson, McCarthy \& Isaacson}{Fischer
  et~al.}{2008}]{Fisher08}
Fischer D.~A.,  Marcy G.~W.,  Butler R.~P.,  Vogt S.~S.,  Laughlin G.,  Henry
  G.~W.,  Abouav D.,  Peek K. M.~G.,  Wright J.~T.,  Johnson J.~A.,  McCarthy
  C.,    Isaacson H.,  2008, ApJ, 675, 790

\bibitem[\protect\citeauthoryear{Frescura, Engelbrecht \& Frank}{Frescura
  et~al.}{2008}]{Frescura08}
Frescura F. A.~M.,  Engelbrecht C.~A.,    Frank B.~S.,  2008, MNRAS, 388, 1693

\bibitem[\protect\citeauthoryear{Gelman}{Gelman}{2008}]{Gelman08}
Gelman A.,  2008, Bayesian Analysis, 3, 445

\bibitem[\protect\citeauthoryear{Gourieroux, Monfort \& Trognon}{Gourieroux
  et~al.}{1984}]{Gourieroux84}
Gourieroux C.,  Monfort A.,    Trognon A.,  1984, Econometrica, 52, 681

\bibitem[\protect\citeauthoryear{Go{\'z}dziewski, Breiter \&
  Borczyk}{Go{\'z}dziewski et~al.}{2008}]{Gozd08}
Go{\'z}dziewski K.,  Breiter S.,    Borczyk W.,  2008, MNRAS, 383, 989

\bibitem[\protect\citeauthoryear{Go{\'z}dziewski, Konacki \&
  Maciejewski}{Go{\'z}dziewski et~al.}{2006}]{Gozd06b}
Go{\'z}dziewski K.,  Konacki M.,    Maciejewski A.~J.,  2006, ApJ, 645, 688

\bibitem[\protect\citeauthoryear{Go\'{z}dziewski, Maciejewski \&
  Migaszewski}{Go\'{z}dziewski et~al.}{2007}]{Gozd07}
Go\'{z}dziewski K.,  Maciejewski A.~J.,    Migaszewski C.,  2007, ApJ, 657, 546

\bibitem[\protect\citeauthoryear{Gregory}{Gregory}{2005}]{Gregory05}
Gregory P.~C.,  2005, in Knuth K.~H.,  Abbas A.~E.,  Morris R.~D.,   Castle
  J.~P.,  eds, Bayesian Inference and Maximum Entropy Methods. Vol.~803 of AIP
  Conf. Proc., A {B}ayesian analysis of extrasolar planet data for {HD}208487.
Am. Inst. Phys., New York, pp 139--145

\bibitem[\protect\citeauthoryear{Gregory}{Gregory}{2007}]{Gregory07a}
Gregory P.~C.,  2007, MNRAS, 374, 1321

\bibitem[\protect\citeauthoryear{Howard, Johnson, Marcy, Fischer, Wright,
  Henry, Isaacson, Valenti, Anderson \& Piskunov}{Howard
  et~al.}{2011}]{Howard11}
Howard A.~W.,  Johnson J.~A.,  Marcy G.~W.,  Fischer D.~A.,  Wright J.~T.,
  Henry G.~W.,  Isaacson H.,  Valenti J.~A.,  Anderson J.,    Piskunov N.~E.,
  2011, ApJ, 726, 73

\bibitem[\protect\citeauthoryear{Lomb}{Lomb}{1976}]{Lomb76}
Lomb N.~R.,  1976, Ap\&SS, 39, 447

\bibitem[\protect\citeauthoryear{Rivera, Laughlin, Butler, Vogt, Haghighipour
  \& Meschiari}{Rivera et~al.}{2010}]{Rivera10}
Rivera E.~J.,  Laughlin G.,  Butler R.,  Vogt S.,  Haghighipour N.,
  Meschiari S.,  2010, ApJ, 719, 890

\bibitem[\protect\citeauthoryear{Robinson, Laughlin, Vogt, Fischer, Butler,
  Marcy, Henry, Driscoll, Takeda \& Johnson}{Robinson
  et~al.}{2007}]{Robinson07}
Robinson S.~E.,  Laughlin G.,  Vogt S.~S.,  Fischer D.~A.,  Butler R.~P.,
  Marcy G.~W.,  Henry G.~W.,  Driscoll P.,  Takeda G.,    Johnson J.~A.,  2007,
  ApJ, 670, 1391

\bibitem[\protect\citeauthoryear{Scargle}{Scargle}{1982}]{Scargle82}
Scargle J.~D.,  1982, ApJ, 263, 835

\bibitem[\protect\citeauthoryear{Vityazev}{Vityazev}{2001}]{Vit-nun}
Vityazev V.~V.,  2001, Analysis of Uneven Time Series.
SPb Univ. Press, Saint Petersburg

\bibitem[\protect\citeauthoryear{Vogt, Butler, Marcy, Fischer, Henry, Laughlin,
  Wright \& A.}{Vogt et~al.}{2005}]{Vogt05}
Vogt S.~S.,  Butler R.~P.,  Marcy G.~W.,  Fischer D.~A.,  Henry G.~W.,
  Laughlin G.,  Wright J.~T.,    A. J.~J.,  2005, ApJ, 632, 638

\bibitem[\protect\citeauthoryear{Vuong}{Vuong}{1989}]{Vuong89}
Vuong Q.~H.,  1989, Econometrica, 57, 307

\bibitem[\protect\citeauthoryear{Zechmeister \& K{\"u}rster}{Zechmeister \&
  K{\"u}rster}{2009}]{ZechKur09}
Zechmeister M.,  K{\"u}rster M.,  2009, A\&A, 496, 577

\end{thebibliography}

\appendix

\section{Meaning of the random timings}
\label{sec_rtime}
It is an essential pecularity of the Vuong test that it treats the timings and measurement
uncertainties as random quantities. This randomness is not related to possible
inaccuracies in the determined values of $t_i$ and $\sigma_i$. Instead, it is related to
the observations scheduling, which is treated as a random process. This issue should be
clarified in more details now.

In the astronomical practice, it is rarely taken into account that the time of an
observation is often a random quantity, broadly analogous to the random measurement error.
It is usually assumed that $t_i$ (as well as $\sigma_i$) are fixed a priori for each data
set analysed. It is assumed that all statistical uncertainties in any data-derived
quantities are caused only by random errors in $x_i$. Basically, we implicitly embed the
observed data set into a general ensemble of hypothetical similar time series, each having
its own sequence of $x_i$ and still the same sequence of $t_i$. This restrictive
assumption about $t_i$ may have undesired effect sometimes. To demonstrate this, assume
that we are carrying out a large survey of many astronomical targets. In practice, the
timings of individual observations are always distributed according to some observational
time window, which often shows some regular patterns (e.g. periodic gaps), but within this
time window they are usually distributed pretty randomly. Usually we cannot take a
snapshot of all targets at once, so each target has individual time series with different
sequence of $t_i$. Under such circumstances, the random scatter of any data-derived
quantity (e.g., some parametric estimation or a test statistic) incorporates an extra
contribution inferred by the random fluctuations of $t_i$. When we subsequently apply our
statistical procedure to each target of such survey, we should encounter, in general, more
blurred uncertainties and more frequent false alarms than we can expect assuming that all
$t_i$ are fixed. Therefore, to deal with this situation properly, we should consider
another general enseble of time series, which admits of random variations in $t_i$ and
$\sigma_i$, as well as in $x_i$.

We do not try to scare the reader by claiming that all results of any data analysis done
so far should be now rechecked taking into account the possible fluctuations of $t_i$. The
quantities used traditionally in the astronomical data analysis, are often statistically
invariable with respect to the exact sequence of $t_i$. ``Statistically invariable'' means
here that the corresponding distribution function is invariable with respect to
fluctuations of $t_i$ within any given distribution pattern, although individual values of
the test statistic usually do depend on $t_i$. For example:
\begin{enumerate}
\item It is well-known that the values of the Lomb-Scargle periodogram are exponentially
distributed, regardless the actual sequence of timings. Consequently, they remain
exponentially distributed for the random $t_i$ too.

\item According to \citet{Baluev08a}, the tail distribution of the maximum peaks of the
Lomb-Scargle periodogram can be approximated by a formula $W e^{-z} \sqrt z$, where $W$ is
proportional to the sample variance of $t_i$. For large $N$, the latter variance is
practically invariable with respect to random fluctuations of $t_i$, if these fluctuations
obey some well-specified time window. Therefore, the distribution of the maximum
periodogram peaks also should not be significantly affected by the fluctuating timings.

\item The maximum-likelihood (or least-square) estimations of various model parameters are
usually asymptotically unbiased for large $N$, with the uncertainty approximately
proportional to $1/\sqrt N$. The distribution of such estimations is asymptotically
invariable with respect to the random variations of $t_i$, if these variations follow any
specified time window. The uncertainties of the estimations may often depend on this time
window, however.
\end{enumerate}

When $t_i$ (and $\sigma_i$) are treated as fixed non-random quantities, the distribution
of the Vuong statistic significantly depends on their exact sequence. This can be easily
demonstrated. We can avoid dealing with any distribution of $t_i$ if we redefine the
$\KLIC$ divergence as
\begin{eqnarray}
\KLIC_{12}'(\btheta_1,\btheta_2,\bmath z) &=& \frac{1}{N}\expect^0_{\bmath x|\bmath z} \sum_{i=1}^N \log\frac{f_1(x_i|z_i,\btheta_1)}{f_2(x_i|z_i,\btheta_2)} = \nonumber\\
&=& \frac{1}{N}\expect^0_{\bmath x|\bmath z} \log\frac{\mathcal L_1(\bmath x|\bmath z,\btheta_1)}{\mathcal L_2(\bmath x|\bmath z,\btheta_2)},
\label{KLIC2}
\end{eqnarray}
where the expression under the $\expect^0$ sign represents now the usual (observed)
log-likelihood ratio for the two models, and the expectation itself is now taken
conditionally to fixed $z_i$. This new divergence measure represents a direct analog of
the quantity~(\ref{KLIC}), but calculated for a particular discrete sequence of $z_i$.
There is no obstacle to use the same Vuong statistic~(\ref{Vuong}) to test the new null
hypothesis $\KLIC_{12}'=0$, which is apparently no worse than testing the hypothesis
$\KLIC_{12}=0$. Indeed, the value of $\KLIC_{12}'$, which represents an average over the
$N$ timings, should converge to $\KLIC_{12}$ for $N\to\infty$. Since we anyway consider
only this asymptotic case, the two measures $\KLIC$ and $\KLIC'$ are just equivalent, and
the quantity $L$ in~(\ref{lmean}) can be used to estimate both. However, the variance of
$L$ is misbehaving: it has systematically different values for fixed $z_i$ and for random
$z_i$. This occures because random fluctuations in $x_i$ and in $z_i$ generate comparable
contributions in the total variance of $L$. Thus the mentioned variance should be
significantly smaller for fixed $z_i$ than for random $z_i$.

It is not hard to show this rigorously. Indeed, when $z_i$ are treated random, all $l_i$
have the same distribution, and it is easy to derive that
\begin{equation}
\expect^0_{\bmath x,\bmath z} v^2 \simeq \disp^0_{x,z} l, \quad
\expect^0_{\bmath x,\bmath z} L = \expect^0_{x,z} l, \quad
N \disp^0_{\bmath x,\bmath z} L = \disp^0_{x,z} l,
\end{equation}
where $\disp$ stands for the variance operator, $\disp x=\expect x^2-(\expect x)^2$. We
can see that in this case the normalised statistic $\mathcal V=L\sqrt N/v$ indeed should
approximately follow the standard normal distribution. When $z_i$ are fixed, the averaging
over $z_i$ in~(\ref{KLIC2}) and any derived formulae can approximate the expectation
$\expect^0_z$. Applying this rule, we may derive
\begin{equation}
\expect^0_{\bmath x|\bmath z} v^2 \simeq \disp^0_{x,z} l, \quad
\expect^0_{\bmath x|\bmath z} L \simeq \expect^0_{x,z} l, \quad
N \disp^0_{\bmath x|\bmath z} L \simeq \expect^0_{\bmath z} \disp^0_{x|z} l.
\end{equation}
We can see that $v^2$ did not attain any significant systematic bias, as well as $L$,
while the variance of $L$ is now different:
\begin{eqnarray}
N \left(\disp^0_{\bmath x,\bmath z} - \disp^0_{\bmath x|\bmath z} \right) L \simeq
\disp^0_{x,z} l - \expect^0_{z} \disp^0_{x|z} l = \nonumber\\
= \expect^0_z (\expect^0_{x|z} l)^2 - (\expect^0_z \expect^0_{x|z} l)^2 = \disp^0_z \expect^0_{x|z} l \geq 0.
\label{ddiff}
\end{eqnarray}

Therefore, if we consider our situation conditionally to fixed $z_i$, we find that the
variance of the Vuong statistic is systematically smaller than unit. This variance deficit
does not tend to zero when $N\to\infty$. Instead, it stabilizes at a constant value
$\disp^0_z \expect^0_{x|z} l / \disp^0_{x,z} l$. It is difficult to apply the Vuong test
in such situation, since then the exact variance of $\mathcal V$ is unknown even for
$N\to\infty$, although we know that it cannot exceed unit. The variance deficit of
$\mathcal V$ disappears only in a very specific case when $\expect^0_{x|z} l$ is constant
in $z$.

However, we must emphasize again that the Vuong test does not request to specify the
distribution of $t_i$ and $\sigma_i$ explicitly. In practice, we should not care about
this distribution at all, and the practical application of the test is easy. In other
words, although the Vuong test is sensitive to the presence of random fluctuations of
individual timings, it is invariable with respect to their \emph{distribution} relflecting
the shape of the time window.

We still can imagine practical cases falling out of the random interpretation of the
timings. For example, when we take a sequence of images of the same field, it makes the
timings non-random -- they appear always the same for all targets of such fixed-field
survey. In this case, however, the Vuong test does not stop working. As we have just
shown, the variance of the Vuong statistic can only decrease when we move from random
$t_i$ to fixed $t_i$, so when we apply it to such fixed-field survey, we still obtain at
least an upper limit on the false alarm probability. This means that if the Vuong test
recommends to retract the null (equivalence) hypothesis at e.g. $99\%$ level, we can
safely retract it even if the sampling patterns are the same for all our targets: the true
confidence level maybe possibly larger than $99\%$, but this only implies even larger
significance. The only negative consequence is that in the case of so specific data
sampling we can distinguish more close alternatives than usually, and the Vuong test does
not catch this opportunity. That is, another test having better sensitivity in this
specific situation may exist, but it is hardly as general as the Vuong one.

\bsp

\label{lastpage}

\end{document}